\documentclass{iopjournal}

\usepackage{amsmath} 
\usepackage{ragged2e} 
\usepackage{natbib}  

\begin{document}

\articletype{Paper} 

\title{Density Limit Experiments and Core-localized Kinetic MHD Activities in HL-2A Ohmic Heating Plasmas}

\author{L. W. Hu$^1$\orcid{0000-0000-0000-0000},W. Chen$^{1*}$\orcid{0000-0000-0000-0000}, P. W. Shi$^1$\orcid{0000-0000-0000-0000}, T. Long$^1$\orcid{0000-0000-0000-0000}, J. Q. Xu$^1$\orcid{0000-0000-0000-0000}, R. R. Ma$^1$\orcid{0000-0000-0000-0000}, Y. G. Li$^1$\orcid{0000-0000-0000-0000}, L. M. Yu$^1$\orcid{0000-0000-0000-0000}, X. Yu$^1$\orcid{0000-0000-0000-0000}, M. Jiang$^1$\orcid{0000-0000-0000-0000}, T. F. Sun$^1$\orcid{0000-0000-0000-0000}, J.M. Gao$^1$\orcid{0000-0000-0000-0000} , Y. B. Dong$^1$\orcid{0000-0000-0000-0000}, X. L. Zhu$^1$\orcid{0000-0000-0000-0000}, Z. B. Shi$^1$\orcid{0000-0000-0000-0000}}

\affil{$^1$Joint Laboratory for Fusion Product and Energetic Particle, Southwestern Institute of Physics, P.O. Box 432 Chengdu 610041, China}


\affil{$^*$W. Chen}

\email{chenw@swip.ac.cn}


\justifying 
\begin{abstract}
 \justifying 
The density limit is a mysterious barrier to magnetic confinement nuclear fusion, and is still an unresolved issue. In this paper, we will present the experimental results of the density limit and core-localized kinetic MHD instabilities on HL-2A. Firstly, the high density shots with $ne/ne_G>1$ have been achieved by the conventional gas-puff fuelling method in Ohmic heating plasmas, and the corresponding duration time is close to $t\sim500$ ms ($\sim$ $30\tau_E$), where $\tau_E$ is the global energy confinement time. Secondly, it is found for the first time that there are kinetic MHD instabilities in the core plasmas while $ne/ne_G\sim1$. The analysis suggests that the core-localized MHD activities belong to Alfv{\'e}nic ion temperature gradient (AITG) modes or kinetic ballooning modes (KBM), and firstly it is found on experiment that they trigger the minor or major disruption of bulk plasmas while the density profile is peaked. These new findings are of great importance to figure out and understand the origin of the density limit.
\end{abstract}

\section{Introduction}
  High plasma density ($ne$) is essential for accessing high fusion gain since the fusion power density ($P$) scales as $P \propto ne^2$. However, there is a limit (known as Greenwald limit) for tokamak high density shots\cite{Greenwald_1988, Greenwald_2002}. The Greenwald limit is an empirical limit for the achievable line-averaged plasma density on experiments, namely $ne_G=I_p/\pi a^2$, where $ne_G$ is the line-averaged plasma density in units of $10^{20}m^{-3}$, $I_p$ the plasma current in $\mathrm{MA}$ and $a$ the minor radius in $\mathrm{m}$. Generally, when the Greenwald density is reached, the bulk plasma frequently disrupts as well as the shot halts. Therefore, the density limit represents an operational limit for tokamaks. It is crucial to realize the steady-state high density operation. For the ITER baseline scenario, the density of H-mode plasmas is up to $0.85ne_G$\cite{Kirneva_2015}. Many experimental results indicate that the density limit occurrence is correlated to the plasma edge cooling\cite{Rapp_1999}, multifaceted asymmetric radiation from edge (MARFE)\cite{Pucella_2013}, current channel shrinkage, macroscopic magnetohydrodynamics (MHD) activities (mainly tearing modes)\cite{White_Gates_Brennan_2015,Teng_Brennan_2016,Long_2021,Long_2024}, edge turbulence\cite{Giacomin_Ricci_2020,Giacomin_Ricci_2022}, and so forth. These results indicate that the density limit originates from the plasma edge region. Meanwhile, some experimental results also suggest the density limit can be exceeded by the plasma core fuelling\cite{Lang_Suttrop_2012}, edge pumping, or modification of particle transport, which leads to peaked density profiles\cite{Bell_1992}. The physical mechanisms governing the density limit of plasmas are not yet fully understood because they involve a complex array of interrelated phenomena.​

Prior to the onset of density limit disruptions, tearing mode instabilities often emerge frequently. The prevailing explanation attributes this phenomenon to an increased current gradient caused by current quenching following edge cooling\cite{Gates_2012}. However, in the absence of tearing modes, this account still fails to clarify the underlying reason for the restriction on core density rise. The transition from linear Ohmic confinement (LOC) to saturated Ohmic confinement (SOC) in Ohmic heated plasmas \cite{Rice_2020} suggests that as the plasma approaches the density limit, certain mechanisms degrade its energy confinement. A plausible explanation for this degradation involves micro-turbulence or various instabilities. Notably, the Alfv{\'e}nic ion temperature gradient (AITG) mode has been identified during the SOC phase in Ohmic heated plasmas on the HL-2A tokamak \cite{Chen_2016_EPL}. Such instabilities may be linked to the deterioration of plasma confinement. Experimentally, AITG can also be referred to as kinetic ballooning modes (KBM). Simulations using various numerical codes indicate that the excitation of the KBM/AITG mode depends on a specific $\beta$ (pressure ratio) threshold and becomes more unstable under weak magnetic shear conditions \cite{Connor_1978, Belli_Candy_2010, Kumar_2021}. Besides the KBM/AITG mode, other instabilities such as the ion temperature gradient (ITG) mode \cite{Xu_Chen_2023} and the micro-tearing mode (MTM) \cite{drake1977kinetic} may also exist in the plasma core. MTM is primarily destabilized by the electron temperature gradient, while the ITG mode is mainly driven by the ion temperature gradient. Moreover, the ITG mode is stabilized in high-beta plasmas. In high-density plasmas produced by Ohmic heating, the parameter $\beta$ exhibits an increase with rising density, and a substantial temperature gradient is not sustained. Thus, it is unlikely that either mode constitutes the primary cause of the observed density limit. In this paper, we provide a detailed description and analysis of experimental results on density limits and MHD activities obtained from the HL-2A tokamak.

  The experiments reported here were conducted on the HL-2A tokamak \cite{duan2009overview} in lower single-null divertor deuterium plasmas, with a plasma current $I_p \simeq 140-160$ $\mathrm{kA}$ and a toroidal magnetic field $B_t \simeq 1.1-1.4$ $\mathrm{T}$. The device has a major radius $R_0 = 1.65$ m and a minor radius $a=0.4$ m, yielding an almost circular poloidal cross-section. The plasma-facing components are primarily made of graphite tiles. The line-averaged electron density is detected using a multi-channel HCOOH laser interferometry, and density profile is reconstructed via Abel inversion\cite{li2017new}. Density fluctuations were monitored non-locally using a two-channel microwave interferometry in the core region\cite{Shi_2016}. Electron temperature profiles were obtained from an optically thick, second-harmonic (X-mode) electron cyclotron emission (ECE) radiometer \cite{shi2014calibration}, while the central electron temperature was measured by Thomson scattering. Magnetic fluctuations were detected using toroidal and poloidal Mirnov coils\cite{Xiaoquan_2006}. To access high-density regimes, continuous gas puffing was employed, successfully producing peaked density profiles with $ne/ne_G>1$，where $ne$ is the line-averaged electron density. The gas puff primarily raised the edge density, and through plasma self-organization, the entire density profile was elevated. The plasma density limit is known to occur across various types of heated plasmas. To minimize the influence of auxiliary heating methods—such as netural beam injection (NBI) and radio-frequency heating—this study focuses primarily on Ohmic heating plasmas in the HL-2A tokamak.

Section 2 presents the phenomenon of density perturbations in plasmas approaching the density limit, including broad-spectrum turbulence and core-localized magnetohydrodynamic (clMHD) instabilities. Section 3 introduces the experimental results from the HL-2A Ohmic heated plasmas, where gas-puffing is used to exceed the Greenwald density limit. Section 4 describes the conditions under which clMHD instabilities are easier to emerge. In section 5, clMHD modes are identified as KBM/AITGs and their impact on the plasma are analyzed. Finally, a summary in section 6.

\section{Density perturbations in plasma close to density limit}
  
  \begin{figure}[!t]
 \centering
        \includegraphics[scale=0.65]{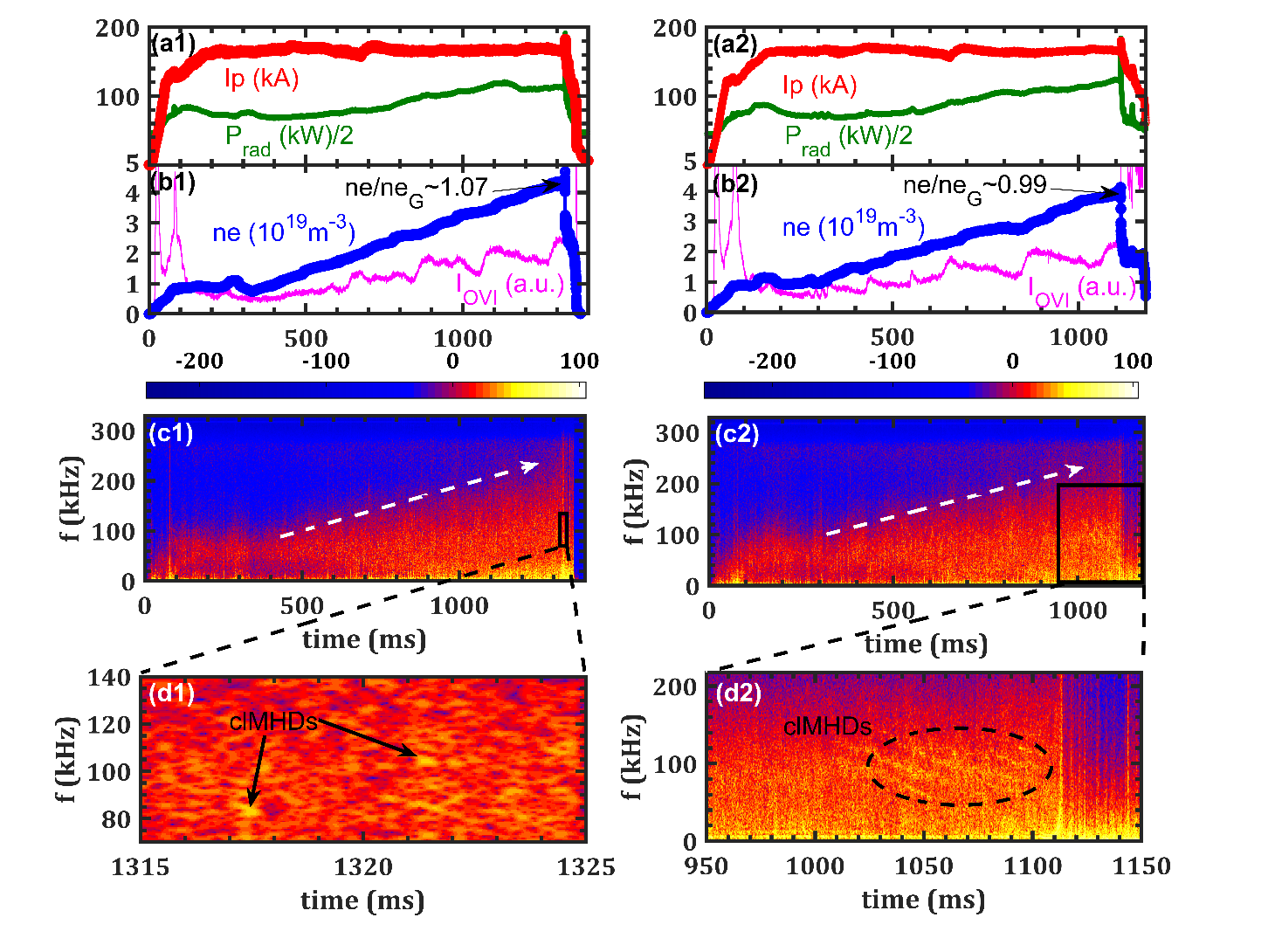}
 \caption{(a1) Time traces of plasma current $I_p$ and total plasma radiation power $P_{rad}$for shot \#38261. The red and green solid lines correspond to $I_p$ and $P_{rad}$, respectively. (a2) Time traces of $I_p$ and $P_{rad}$ for shot \#38262. (b1) Time traces of line-averaged plasma density $ne$ and oxygen impurity spectral line $I_{OVI}$ for shot \#38261. The blue and magenta solid lines show $ne$ and $I_{OVI}$, respectively. The black arrow marks the relative density level at the time of peak density. (b2) Time traces of $ne$ and $I_{OVI}$ for shot \#38262. (c1) Spectrum of microwave interferometry signal for shot \#38261, where the microwave interferometry channel passes through the center of the plasma. The white arrow indicates the trend of spectral broadening. (c2) Spectrum of the central microwave interferometry signal for shot \#38262.(d1) Magnified view of spectrum as indicated by black rectangular in subfigure (c1). The black arrows indicate the onset time of clMHD modes. (d2) Magnified view of spectrum as indicated by black rectangular in subfigure (c2). The dashed elliptical line denotes signature of the clMHD modes.}
\label{fig1}
\end{figure}

  \begin{figure}[!t]
 \centering
        \includegraphics[scale=0.07]{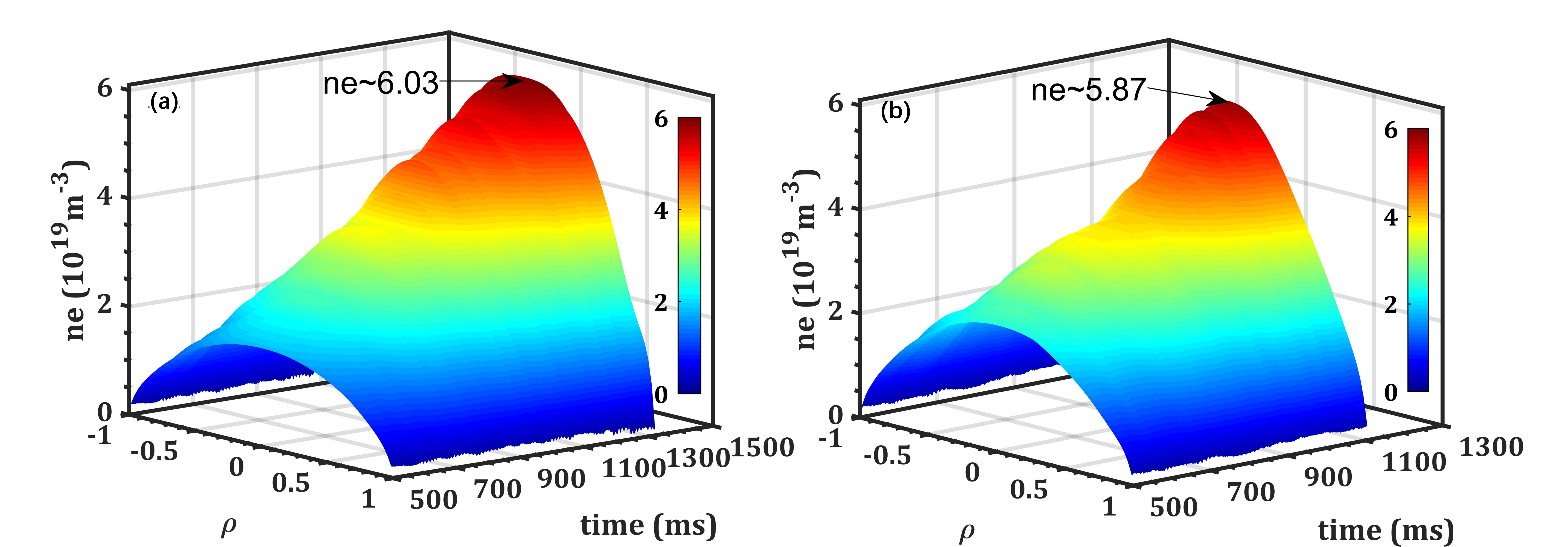}
 \caption{(a) Time evolution of the density profile for shot \#38261, obtained from Abel inversion of laser interferometry data. The black arrow marks the core density before the density-limit disruption, with $ne\sim 6.03\times 10^{19} \mathrm{m^{-3}}$.(b) Time evolution of the density profile for shot \#38262. The core density prior to disruption is $ne\sim 5.87\times 10^{19} \mathrm{m^{-3}}$. Here $\rho=r/a$ is the normalized radius. The observed density ramp-up and peaking in both shots result from plasma self-organization under gas puffing.  }
\label{fig2}
\end{figure}

 Figure \ref{fig1} shows two Ohmic-heated shots in HL-2A tokamak. Shot \#38261 reaches a relative density of $ne/ne_G\sim 1.07$ before disruption, while shot \#38262 attains $ne/ne_G\sim 0.99$, as marked by the black arrows in Figure \ref{fig1}(b1,b2).  The high density is achieved by conventional gas-puff fueling. As shown in Figure \ref{fig1}(a1,a2), the plasma current remains nearly constant during the density ramp-up phase. The total radiation power of plasma $P_{rad}$ increases slightly as the plasma approaches the Greenwald density limit, suggesting that the disruption may not be caused by a radiative collapse. The magenta curves in Figure \ref{fig1}(b1,b2) indicate that the oxygen impurity levels $I_{OVI}$ rises slightly with increasing density, consistent with $P_{rad}$. As indicated by the white arrows in Figure \ref{fig1}(c1,c2), the spectral range of density perturbations observed via microwave interferometry broadens with increasing density, evolving to a maximum frequency of approximately 200 kHz near the density limit. These perturbations are generally considered to be caused by turbulence, as turbulence-induced perturbations typically manifest as broad-spectrum disturbances in the frequency domain. This paper does not delve into the specific types of broad-spectrum turbulence discussed here, and as will be evident from Figure \ref{fig5} later, these broad-spectrum turbulences are not always present in Ohmic heating density limit plasmas. Figure \ref{fig1}(d1, d2) shows the magnified views of the spectra corresponding to the black rectangular areas in Figure \ref{fig1}(c1, c2). The black arrows in Figure \ref{fig1}(d1) denote the onset time of the clMHD instabilities. The dashed elliptical line in Figure \ref{fig1}(d2) indicates the signature of the clMHD modes in the microwave interferometry spectrum. The term ‘clMHD’ is used for the modes observed in Figure \ref{fig1}(d1, d2) for the following reasons: 1. These modes are detectable in the microwave interferometry spectrum but not in the magnetic probe spectrum, indicating that they are localized in the plasma core. 2. Unlike turbulence, these modes exhibit coherent structures characteristic of MHD behavior. It is evident that clMHD modes are present in both shots; however, they exhibit higher intensity in shot \#38262. Since both shots eventually reach the density limit, these clMHD modes are likely not the direct cause of disruption. However, their presence seems to reduce the achievable density: the plasma reaches $ne/ne_G\sim 1.07$ in the case of weak instabilities, but only $ne/ne_G\sim 0.99$ when the instabilities are stronger. This indicates that the clMHD modes lower the maximum attainable relative density.
  
 According to Figure \ref{fig1}(d2), shot \#38262 displays coherent instabilities in the frequency range of 50–150 kHz. The spectra correspond to the interferometry channel passing through the plasma center (z = 0). Notably, these instabilities are absent in the horizontal interferometry channel located at z = 15 cm (where z denotes the longitudinal direction of the device), indicating that the modes are localized within the region where r < 15 cm, with r representing the minor radius. The frequency of a toroidal Alfvén eigenmode (TAE) is approximately given by $f_{TAE}\sim v_A/(4\pi qR)$, where $q$ is the safety factor and $v_A$ is the Alfv$\acute{e}$n velocity and $R$ is the large radius\cite{Nazikian_1998}. At the $q$=1 surface, $f_{TAE}$ is approximately 210 kHz. The instability frequency observed in shot \#38262 is about $\frac{1}{2}f_{TAE}$, suggesting a possible connection with shear Alfvén waves. As shown in Figure \ref{fig1}(d2), the frequency of the instabilities decreases as the plasma density increases. The observed frequency is influenced by the Doppler shift, and can be expressed as $f_{lab}=f_{plasma}\pm nf_{v\phi}$ ,where $f_{lab}$ is the frequency in laboratory frame and $f_{plasma}$ is frequency in plasma frame, $n$ is the toroidal mode number and $f_{v\phi}$ is the toroidal rotation frequency\cite{strait1994doppler}. According to Figure \ref{fig1}(a2,b2) the plasma parameters in shot \#38262 do not change significantly between 1000 ms and 1100 ms. Thus, the decrease in instability frequency may be attributed to a reduction in the toroidal mode number. These instabilities emerge tens of milliseconds before the density-limit disruption and may be partly linked to the onset of the disruption.

  To clarify why instabilities are weaker in shot \#38261 but stronger in \#38262, we analyze in detail the evolution of the density profiles in both shots.  Figure \ref{fig2} shows the time-dependent density profiles for shots \#38261 and \#38262. Both shots exhibit clear density peaking. The peak electron density in shot \#38261 reaches about $\sim 6.03\times 10^{19} \mathrm{m^{-3}}$, while in shot \#38262 it is approximately $ 5.87\times 10^{19} \mathrm{m^{-3}}$. Moreover, the density peaking phenomenon is more pronounced in shot \#38262 than in shot \#38261. A more peaked density profile implies a stronger density gradient. Since both shots are Ohmically heated, a large temperature gradient is unlikely to develop. Therefore, the instabilities observed in shot \#38262 are likely driven by the density gradient of thermal particles. These instabilities may exhibit a threshold in density gradient beyond which they become unstable. Under the present Ohmic heating and gas-puffing conditions, density peaking reflects a self-organizing behavior of the plasma.

\section{Exceeding Greenwald density limit by gas-puffing on HL-2A}
\begin{figure}[!t]
 \centering
        \includegraphics[scale=0.55]{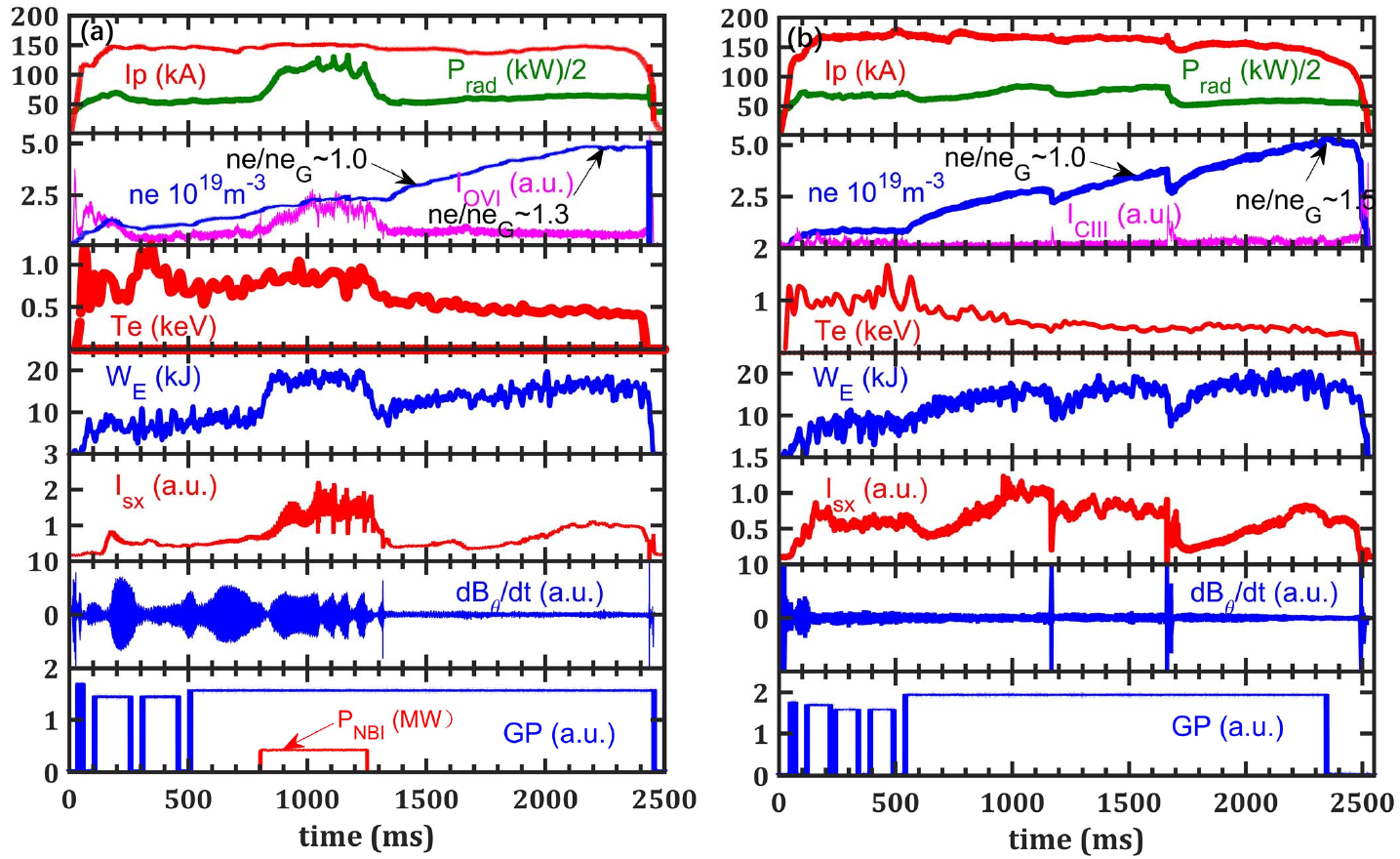}
 \caption{(a) Time traces of plasma parameters for shot \#38522. (b) Time traces of plasma parameters for shot \#38581.The following quantities are shown: plasma current $Ip$ (red line); total radiation power of plasma $P_{rad}$ (green line); line-averaged electron density $ne$ (blue line); oxygen impurity level $I_{OVI}$ (magenta line); carbon impurity level $I_{CIII}$ (magenta line); core electron temperature $Te$ (red line); plasma stored energy $W_E$ (blue line); soft X-ray (SXR) signal $I_{SX}$ (red line); time-differential poloidal magnetic perturbation $dB_{\theta}/dt$ (blue line); gas-puffing signal $GP$ (blue line) and NBI power $P_{NBI}$ (red line). Black arrows mark the times when the ratio $ne/neG$ reaches 1.0 and its maximum value.}
\label{fig3}
\end{figure}

Through gas puffing, a high-density operational regime has been achieved on the HL-2A tokamak, in which the plasma density exceeds the Greenwald density limit. Figure \ref{fig3} shows the time traces of plasma parameters for shots \#38522 and \#38581. In shot \#38522, about 0.5 MW of NBI power is applied from 800 ms to 1250 ms; during other periods, both shots are sustained by Ohmic heating alone. The plasma current is $Ip\sim$ 150 kA in both shots. And the corresponding Greenwald density limit $ne_G$ is about $3.0\times 10^{19} \mathrm{m^{-3}}$. Both shots reach a maximum line-averaged electron density of approximately $5.0\times 10^{19} \mathrm{m^{-3}}$. In shot \#38581, the plasma current gradually decreases as density rises in the later stage, leading to a reduction in the Greenwald limit. For shot \#38522, the maximum ratio of the electron density to the Greenwald density limit, $ne/neG$, is approximately 1.3, while for shot \#38581, it is about 1.5. In both shots, the plasma remains above the Greenwald limit for more than 500 ms—roughly 30 times the energy confinement time $\tau_E$—indicating that the Greenwald limit does not constitute a strict barrier to plasma density. The ability to attain such high $ne/neG$ values may be attributed to relatively low impurity content and comparatively weak plasma instabilities. As shown in the time traces of impurity signals ($I_{OVI}$ and $I_{CIII}$ ) and radiation power ($P_{rad}$) in Figure \ref{fig3}, neither the overall radiation nor the impurity levels increase significantly before the density-limit disruption. During NBI heating in shot \#38522, a slight rise in radiated power and oxygen impurity is observed, likely due to confinement degradation caused by the NBI. Tearing modes are detected in the magnetic signals of shot \#38522 before 1300 ms, but they are weak until the density-limit disruption happens at about 2500 ms.

Shot \#38581 experiences two minor disruptions, identified by a sudden increase in the amplitude of the poloidal magnetic perturbation signal, $dB_{\theta}/dt$. At the time of these minor disruptions, the plasma density is close to the Greenwald limit, with $ne/ne_G\sim 1.0$. In both discharges, the electron temperature decreases as density increases due to gas puffing. As a result, the plasma stored energy $W_E$ does not rise significantly at high density, consistent with the scaling $W_E\propto ne\cdot Te$. The decrease in plasma temperature following density increase via fueling can be attributed to the following factors: 1. Direct cooling from the injection of low-temperature gas; 2. Increased radiation losses;  As previously noted, the stored energy of the injected gas is negligible compared to that of the plasma. Thus, gas injection does not contribute to an increase in plasma stored energy, in accordance with energy conservation. Therefore, the nearly constant stored energy together with the gradually increasing density corresponds to a decreasing plasma temperature. This process describes the mechanism of direct cooling of the plasma by low-temperature gas. The radiation power of plasma scales as $S_B\sim Z_{eff}n^2Te^{1/2}$, where $Z_{eff}$ is the effective charge number\cite{freidberg2008plasma}. Although impurity levels often rise at high density due to degraded confinement—leading to an increase in $Z_{eff}$—the radiated power shown in Figure \ref{fig3} remains nearly constant. This indicates that increased radiation is not the primary cause of the temperature decrease in high-density plasmas.

 \begin{figure}[!htbp]
 \centering
        \includegraphics[scale=0.07]{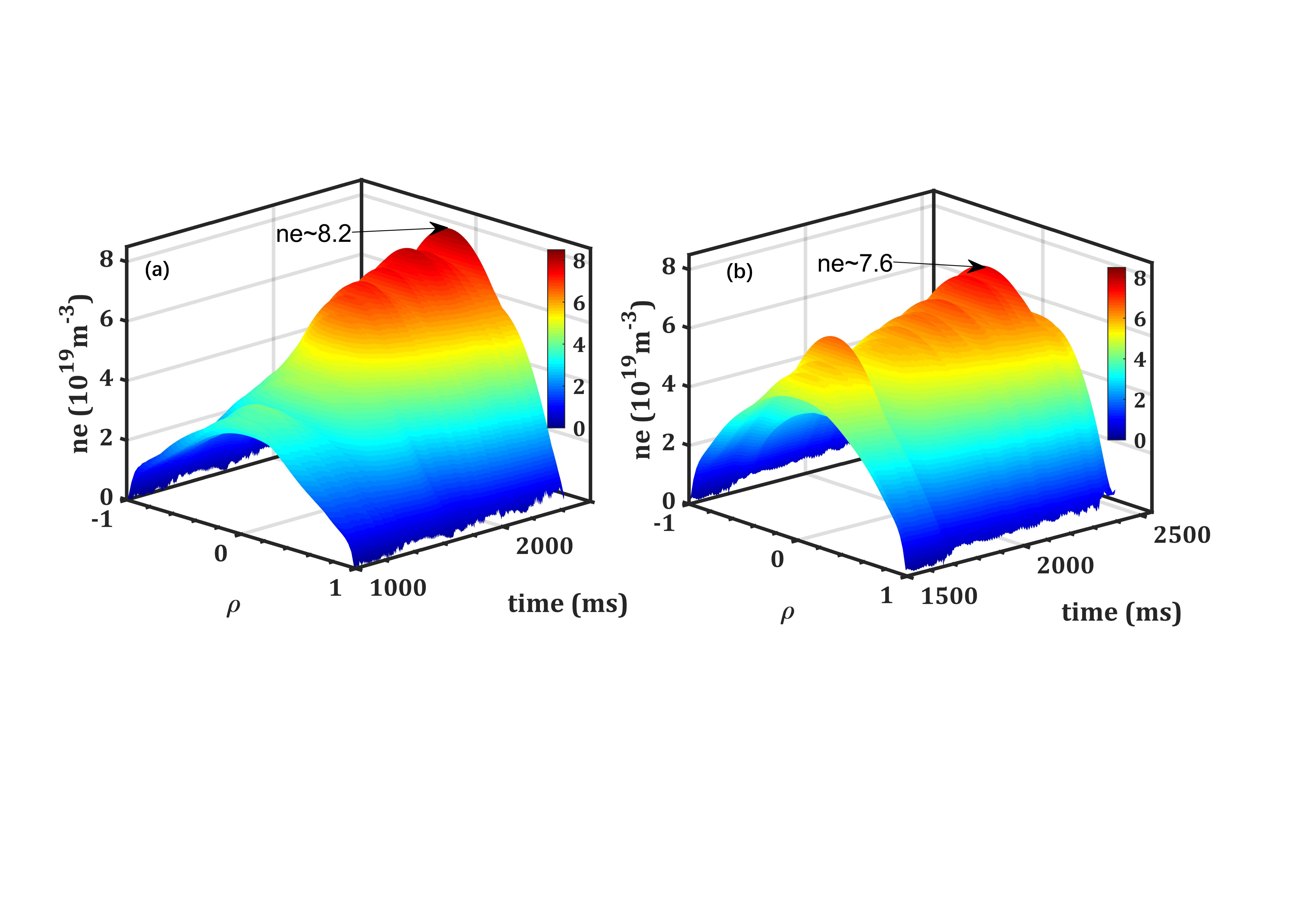}
 \caption{(a) Time evolution of the density profile for shot \#38522, obtained from Abel inversion of laser interferometry data. The black arrow indicates the core density before the density-limit disruption, $ne\sim 8.2 \times 10^{19} \mathrm{m^{-3}}$. (b) Time evolution of the density profile for shot \#38581. The core density prior to disruption is $ne\sim 7.6\times 10^{19} \mathrm{m^{-3}}$. Here $\rho=r/a$ denotes the normalized radius. The overall drop in plasma density after 1500 ms corresponds to a minor disruption occurring around 1660 ms.}
\label{fig4}
\end{figure}

 \begin{figure}[!t]
 \centering
        \includegraphics[scale=0.9]{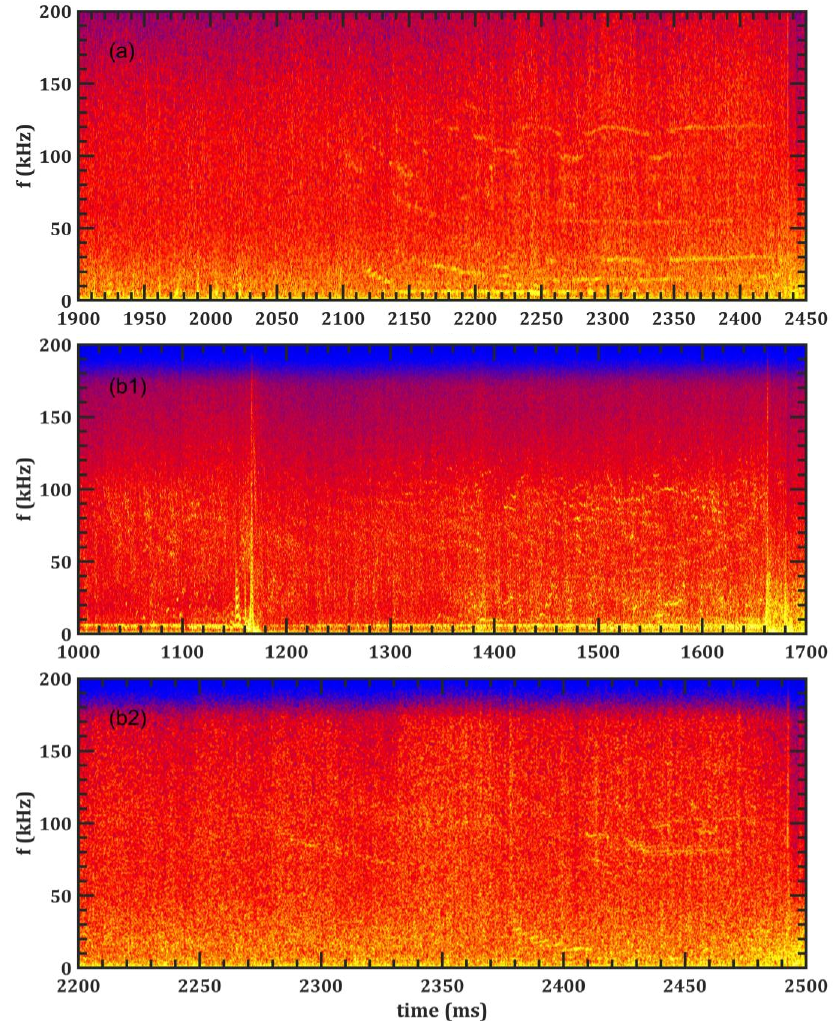}
 \caption{(a) Microwave interferometry spectrum before the density-limit disruption in shot \#38522, showing instabilities in the 0–130 kHz range.(b1) Microwave interferometry spectrum before the minor disruption in shot \#38581, with instabilities emerging in the 0–100 kHz range.(b2) Microwave interferometry spectrum before the major density-limit disruption in shot \#38581.}
\label{fig5}
\end{figure}

Figure \ref{fig5}(a) shows the microwave interferometry spectrum for shot \#38522. In this discharge, the electron density exceeds the Greenwald limit after 1900 ms. After 2100 ms, multiple instabilities appear in the spectrum, with frequencies ranging from near zero up to 150 kHz. By combining Figure \ref{fig1}(d2) and Figure \ref{fig5}(a), it can be seen that these clMHD modes do not necessarily occur at a specific $ne/neG$ value (e.g., $ne/neG\sim 1$), but rather emerge tens of milliseconds before a density limit disruption. This experimental observation suggests that these clMHD modes may play a significant role in the density limit disruption process. For instance, in shot \#38262, clMHD modes appear just before $ne/neG$ reaches $\sim 1.0$, followed shortly by a density limit disruption. In contrast, in shot \#38522, although the plasma density has already exceeded the Greenwald density limit prior to 2100 ms, a disruption does not occur immediately. It is only after 2100 ms, with the continued density increase and further evolution of the clMHD modes, that the disruption ultimately takes place.

 The TAE frequency at $q$=2 surface is about 96 kHz during the time period depicted in Figure \ref{fig5}(a). Based on frequency, the clMHD modes in the spectrum are more consistent with instabilities such as KBM (also referred to as AITG) or beta-induced Alfvén eigenmode (BAE) \cite{Chen_2016,Chen_2018}. It should be noted that for the KBM/AITG, the frequency satisfies $f_{KBM}\sim f_{BAE}<f_{TAE}$. Moreover, $f_{KBM}\sim f_{*pi}\sim f_{ti}$, where $f_{*pi}$ is the ion diamagnetic drift frequency and $f_{ti}$ is the ion transit frequency.\cite{Zonca_Chen_Santoro_1996}. The ratio $f_{BAE}/f_{TAE}$ is given approximately by $\sim 2q\sqrt{(7/4)\beta_i+\beta_e}$. For shot \#38522, using typical core values $q\sim 2$ and $\beta=\beta_i+\beta_e\sim 0.005$, this ratio is about 0.47. Both experimental and theoretical studies suggest that KBM/AITG frequencies are generally lower than those of BAE, and their durations are shorter \cite{ma2023low}. This is because KBM/AITG excitation is highly sensitive to the presence of rational surfaces, whereas BAE is less sensitive. The instabilities below 30 kHz, appearing as “Christmas lights” in the spectrum, are likely KBM/AITG modes. Since TAE excitation requires a particle energy threshold that is generally not met in Ohmic plasmas due to the lack of energetic particles \cite{Hou_2018}, the TAE instabilities are not considered in this paper.

Figure \ref{fig5}(b1) shows the microwave interferometry spectrum for shot \#38581 during the interval 1000–1700 ms. Two minor disruptions are observed in this shot, appearing in the spectrum as rapidly growing disturbances emerging from near zero frequency. Shortly before each minor disruption (at approximately 1160 ms and 1680 ms), multiple instabilities below 100 kHz also arise. The spectrum again shows instabilities that can be grouped into two frequency categories. The TAE frequency at $q$=2 surface is about 120 kHz during the shown time period in figure \ref{fig5}(b1). The frequency of instabilities in figure \ref{fig5}(b1) is in the range of $(0,f_{TAE})$. Figure \ref{fig5}(b2) displays the spectrum from 2200 ms to 2500 ms for the same shot. Similar low-frequency “Christmas light”-like instabilities and high-frequency instabilities also appear prior to the major disruption.

As shown in Figure \ref{fig1}(b1,b2), the line-averaged plasma density in shots \#38261 and \#38262 rises continuously before density-limit disruptions occur. In contrast, for shot \#38522 under continuous gas puffing (Figure \ref{fig3}(a)), the line-averaged density stops increasing once it reaches its maximum (corresponding to $ne/neG \sim$ 1.3). Two possible explanations can be considered for this behavior. One is that the high plasma density degrades the fueling efficiency of gas puffing, preventing further density increase. The other is that after the peak density is reached, some plasma-internal mechanism—such as specific instabilities—limits further density rise.

The first explanation is plausible because, as indicated in Figure \ref{fig4}, the peak core density in shot \#38522 is about 8.2 $\times 10^{19} \mathrm{m^{-3}}$,while in shot \#38581 it is approximately 7.6 $\times 10^{19} \mathrm{m^{-3}}$. Despite shot \#38581 having a higher relative density $ne/neG$ than shot \#38522 (Figure \ref{fig3}), its core density is lower. This apparent contradiction stems from the gradual decrease in plasma current in shot \#38581, which reduces its Greenwald limit $neG$, while the current in shot \#38522 remains steady. Moreover, although shot \#38581 reaches a higher line-averaged density, its core density is lower, implying a flatter density profile with a smaller overall gradient, as seen in Figure \ref{fig4}. This may explain why more instabilities are observed in the microwave interferometry spectrum of shot \#38522 than in that of shot \#38581 (Figure \ref{fig5}(a, b2)).

 The second explanation also remains viable. According to Figure \ref{fig3}(a), the line-averaged density in shot \#38522 stops increasing after about 2100 ms, even though gas puffing continues until approximately 2450 ms. Coincidentally, a variety of instabilities emerge in the core region starting around 2100 ms (Figure \ref{fig5}(a)), suggesting that these instabilities could indeed be responsible for limiting further density increase.

\section{​Scenario analysis for the onset of clMHD modes​}

\begin{figure}[!h]
 \centering
        \includegraphics[scale=0.5]{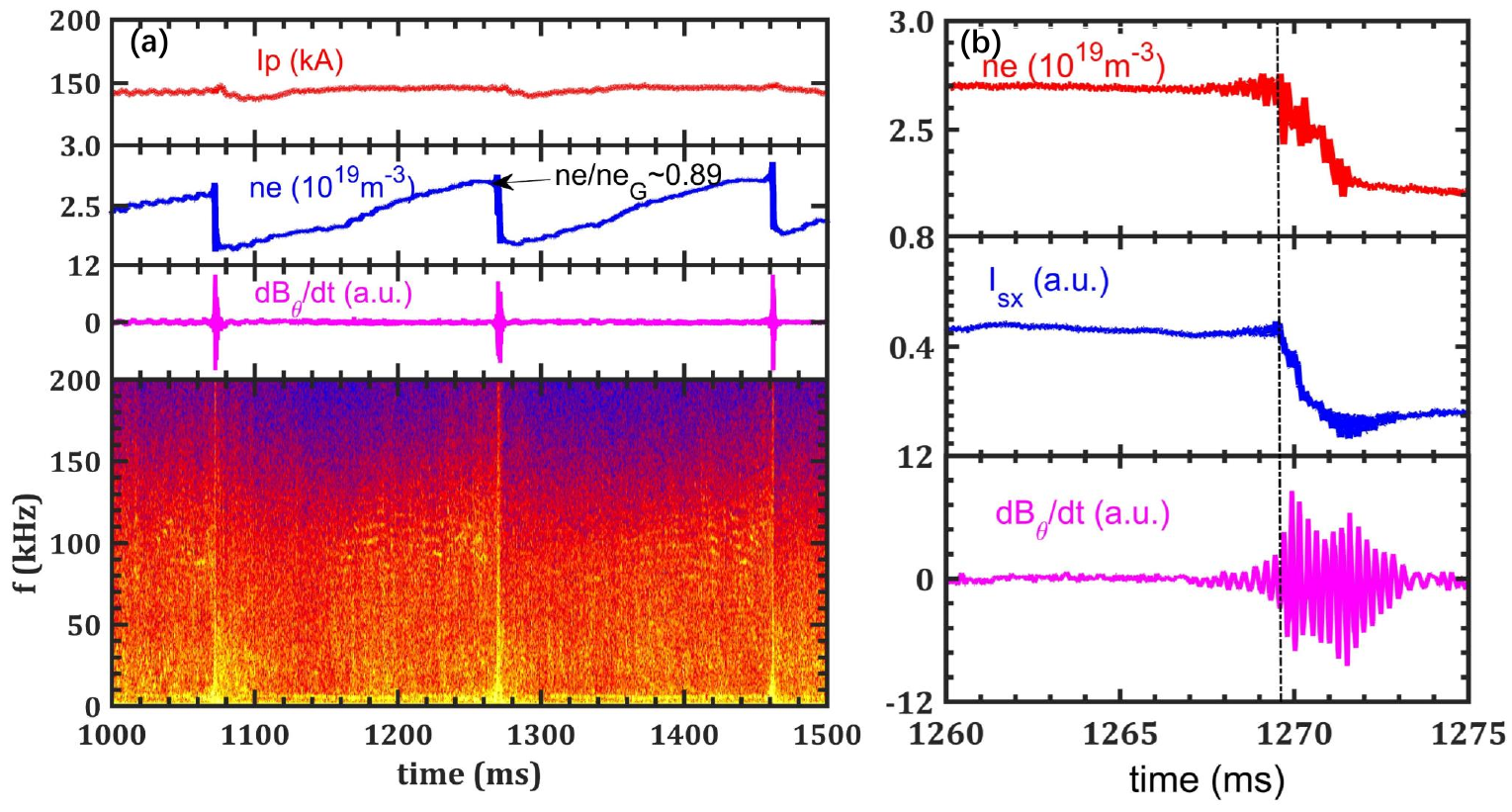}
 \caption{(a) Time traces of plasma parameters and microwave interferometry spectrum for shot \#38570. Black arrows mark the ratio $ne/neG$ prior to each density-limit minor disruption. Instabilities are observed in the interferometry spectrum before all three minor disruptions shown. (b)Zoomed-in view of plasma parameters for the same shot. The vertical dashed line indicates the time of a minor disruption. Displayed traces include: plasma current $Ip$ (red), electron density $ne$ (blue), time-differential poloidal magnetic perturbation $dB_{\theta}/dt$ (magenta) and soft X-ray level $I_{SX}$ (blue).}
\label{fig6}
\end{figure}

Figure \ref{fig6} shows the time traces of plasma parameters and the microwave interferometry spectrum for shot \#38570. In this discharge, three minor disruptions occur between 1000 ms and 1500 ms. Minor disruptions are identified by small upward spikes in the plasma current, accompanied by a rapid drop in plasma density or a sharp rise in magnetic perturbations. Since the maximum Greenwald fraction reaches $ne/ne_G\sim0.89$, the minor disruptions are attributed to the density limit. In the tens of milliseconds preceding each minor disruption, instabilities in the frequency range of about 70–120 kHz repeatedly appear in the microwave interferometry spectrum. Zoomed-in views of the plasma parameters are provided in Figure \ref{fig6}(b). The vertical dashed line marks the time of a minor disruption. Following the occurrence of a minor disruption, the SXR level and magnetic perturbation begin to increase. Notably, the density perturbations already rise before the disruption, indicating that they act as precursors to the minor density-limit disruptions. ​As shown in Figures \ref{fig5} and \ref{fig6}, when the plasma density approaches the density limit, the coincidental occurrence of clMHD modes makes the plasma more likely to experience either a major or minor disruption.​ Corresponding to the recurrent instances of clMHD modes and minor disruptions near the density limit in Figure \ref{fig6}(a), the physical process can be interpreted as a cyclic sequence: the density increases toward the Greenwald limit, triggering the onset of clMHD modes, which leads to a minor disruption. This event causes the density to decrease, resulting in the cessation of clMHD modes, after which the density begins to rise again. The repeated occurrence of such clMHD modes in a single discharge implies that their appearance is facilitated by a higher plasma density or a more peaked density profile.

 \begin{figure}[!htbp]
 \centering
        \includegraphics[scale=0.5]{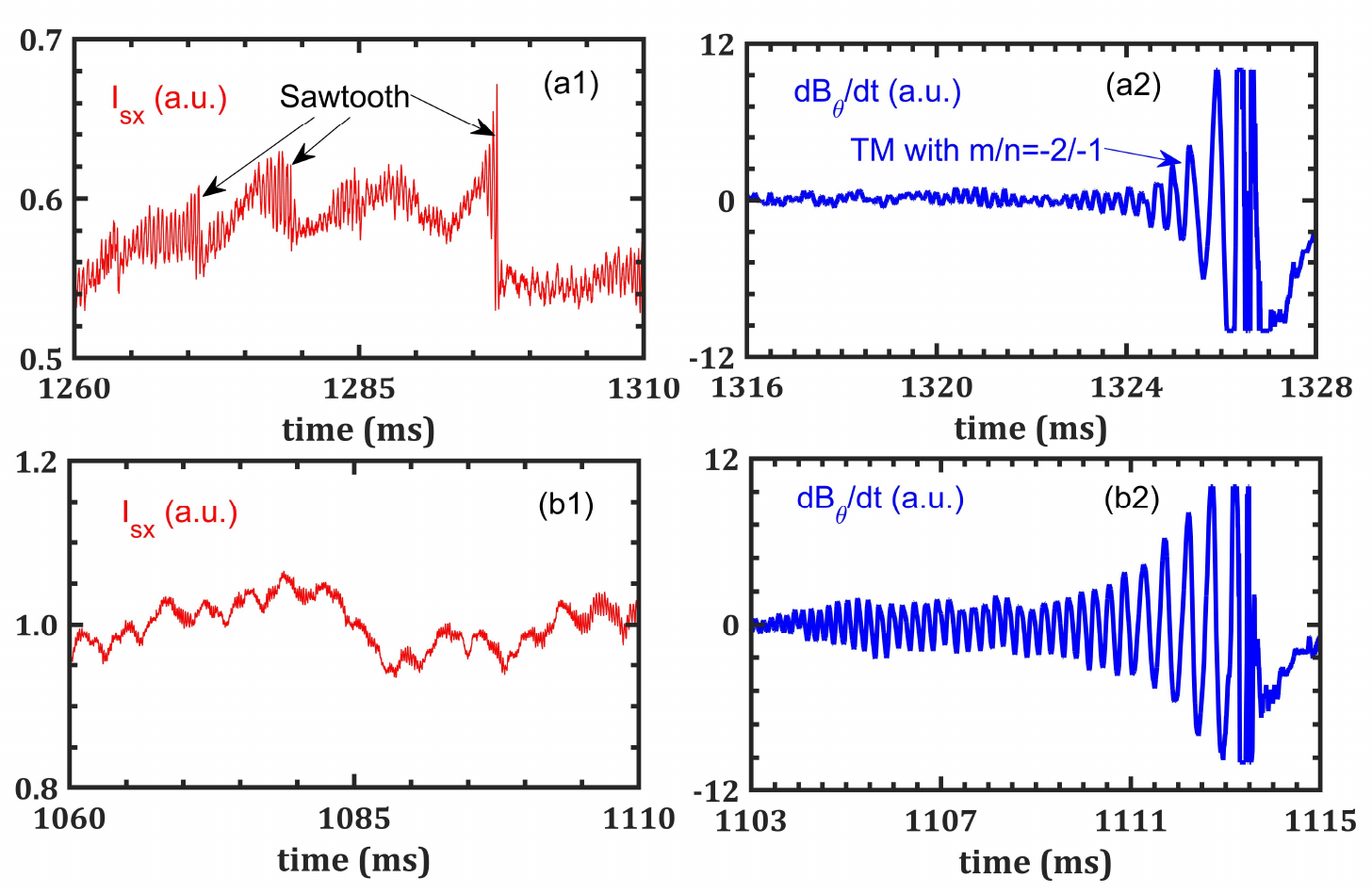}
 \caption{(a1) Raw SXR signal $I_{SX}$ (red) for shot \#38261. The black arrows indicate the moments of sawtooth crashes. (a2) Raw signal of time-differential poloidal magnetic perturbation $dB_{\theta}/dt$ (blue) for shot \#38261. The blue arrow indicates the occurrence of a tearing mode (TM) with mode numbers m/n=-2/-1. (b1) Raw SXR signal $I_{SX}$ (red) for shot \#38262. (b2) Raw signal of time-differential poloidal magnetic perturbation $dB_{\theta}/dt$ (blue) for shot \#38262.}
\label{fig7}
\end{figure}

Figure \ref{fig7}(a1) shows the raw SXR signal for shot \#38261, with sawtooth oscillations visible between 1260 ms and 1310 ms. ​This indicates that the safety factor in the plasma core is less than unity ($q$ < 1), hence the presence of a $q$ = 1 surface.​ Figure \ref{fig7}(a2) presents the raw time derivative of the poloidal magnetic perturbation, $dB_{\theta}/dt$, for the same shot. A tearing mode (TM) with mode numbers m/n=-2/-1 appears after 1324 ms, ultimately leading to a density-limit disruption. Figure \ref{fig7}(b1) displays the raw SXR signal for shot \#38262. Here, the SXR intensity begins to decrease around 1080 ms, and no sawtooth activity is observed. Figure \ref{fig7}(b2) shows the corresponding $dB_{\theta}/dt$ signal for shot \#38262, where a TM also causes the disruption. Compared to shot \#38261, however, the $dB_{\theta}/dt$ signal in shot \#38262 exhibits a larger amplitude even before the TM emerges.

 As seen in Figure \ref{fig1}, clMHD modes are present in the density perturbation spectrum prior to the disruption. However, in shot \#38261, the clMHD modes are weaker in intensity and fewer in number, corresponding to the scenario shown in Figure \ref{fig7}(a1, a2). In shot \#38262, the clMHD modes exhibited greater intensity and more in number, aligning with the case depicted in Figure \ref{fig7}(b1, b2). Thus, clMHD modes are more likely to be excited when the core safety factor $q_0$ exceeds 1.

 \begin{figure}[!htbp]
 \centering
        \includegraphics[scale=0.5]{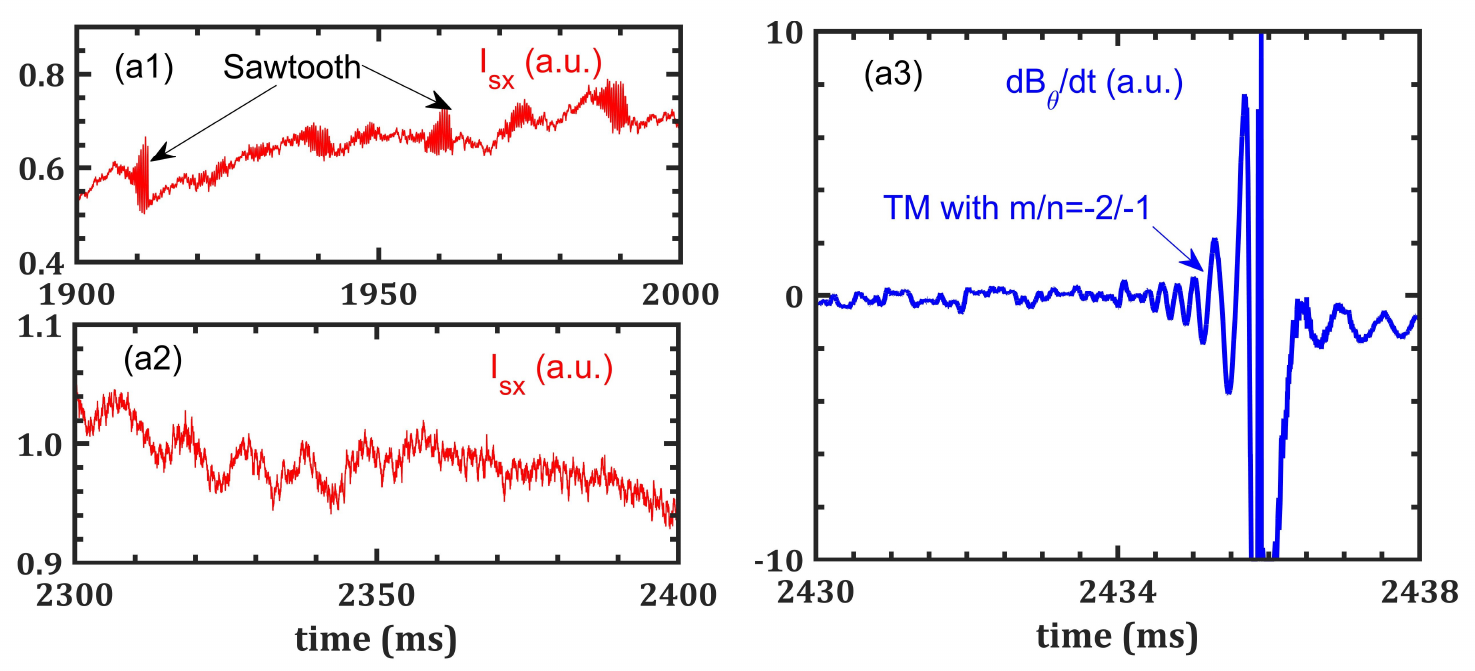}
 \caption{(a1) Raw SXR signal $I_{SX}$ for shot \#38522. Black arrows mark sawtooth crashes. (a2) Raw signal of SXR level $I_{SX}$ for shot \#38522. (a3) Raw time derivative of poloidal magnetic perturbation $dB_{\theta}/dt$ for shot \#38522. Blue arrow indicates the occurrence of a TM with mode numbers m/n=-2/-1.}
\label{fig8}
\end{figure}

 \begin{figure}[!h]
 \centering
        \includegraphics[scale=0.35]{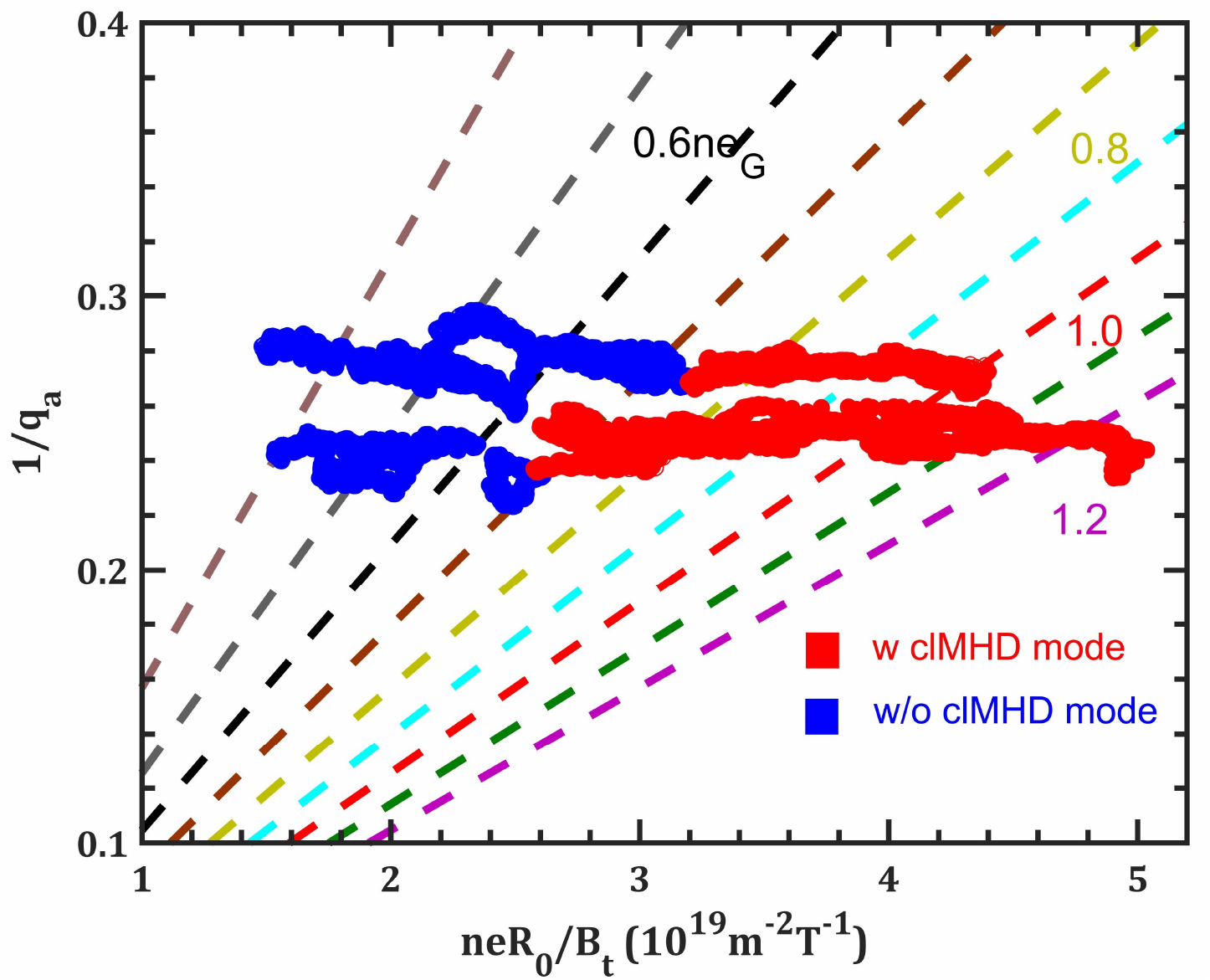}
 \caption{Hugill plot for high-density ohmic-heating plasmas in HL-2A. The safety factor at the plasma boundary is denoted by $q_a$. The dashed lines in plot indicates relationship with Greenwald density limit. Red and blue symbols represent discharges with and without clMHD modes, respectively. Here, only shots where clMHD modes occur are counted.}
\label{fig9}
\end{figure}

In a single shot, a similar phenomenon is observed before and after the appearance of clMHD modes. Figures \ref{fig8}(a1) and (a2) present the raw SXR signals obtained during shot \#38522 show the raw SXR signals from shot \#38522 during two different time intervals: one without clMHD modes in the microwave interferometry spectrum (a1), and the other with clMHD modes present (a2). The appearance of sawtooth oscillations in (a1) indicates the existence of a rational surface with safety factor $q$=1 at that time. In this interval, the SXR intensity $I_{SX}$ rises over time, consistent with the increasing plasma density. In contrast, Figure \ref{fig8}(a2) shows a gradual decline in $I_{SX}$ over time, even though the plasma density continues to increase (as seen in Fig. \ref{fig4}). This decrease in SXR intensity can be attributed to a drop in electron temperature. Figure \ref{fig8}(a3) further shows that a rapidly growing m/n=-2/-1 TM triggers a density-limit disruption. Integrating these results with those from Fig. \ref{fig7}, it can be inferred that the safety factor profile—or more broadly, the magnetic configuration of the plasma—significantly influences the clMHD instabilities that precede the density limit.

The physical scenario of the density limit in HL-2A ohmic-heated plasmas can be summarized as follows. Following gas puffing, the plasma density rises and the density profile becomes more peaked, accompanied by an overall decrease in electron temperature. As the density approaches the Greenwald limit, strong density gradient or $\beta$-induced clMHD modes develop in the plasma. These clMHD modes generally appear as density perturbations. Typically, within tens of milliseconds after the appearance of these clMHD modes, tearing modes exhibit rapid growth, leading to plasma disruption. The impact of clMHD modes on tearing modes requires further investigation, such as their effects on transport and current density profiles.

Figure \ref{fig9} shows the Hugill plot \cite{Stabler_1992} for high-density Ohmic-heated plasmas in the HL-2A tokamak. Red symbols correspond to discharges with clMHD modes, while blue symbols represent those without such modes. A clear threshold is observed for the occurrence of these clMHD modes at approximately $0.7 ne_G$ in HL-2A Ohmic-heated plasmas.  As shown in Figure \ref{fig9}, these clMHD modes occur within the range 0.23 < $1/q_a$ < 0.28, corresponding to 3.6 < $q_a$ < 4.3. In addition, as indicated in Figure \ref{fig7} and \ref{fig8}, these clMHD modes often arise in the absence of sawtooth oscillations, implying that the core safety factor $q_0$ exceeds 1 when these modes are present. At the same time, since the eventual density-limit disruption is triggered by m/n=-2/-1 TMs, the core safety factor $q_0$ should be less than 2. Therefore, the safety factor profile can be estimated as follows: in the core region,1 < $q_0$ < 2, and at the plasma edge, 3.6 < $q_a$ < 4.3.  Due to the absence of auxiliary heating, the safety factor increases slowly and monotonically from the core to the edge, resulting in weak magnetic shear $s=(r/q)(dq/dr)$ in the plasma core. This suggests that these clMHD modes are more easily excited under conditions of weak magnetic shear and high plasma density.

\section{​Identification of clMHD modes as KBM/AITG and their impact​}

 \begin{figure}[!h]
 \centering
        \includegraphics[scale=0.5]{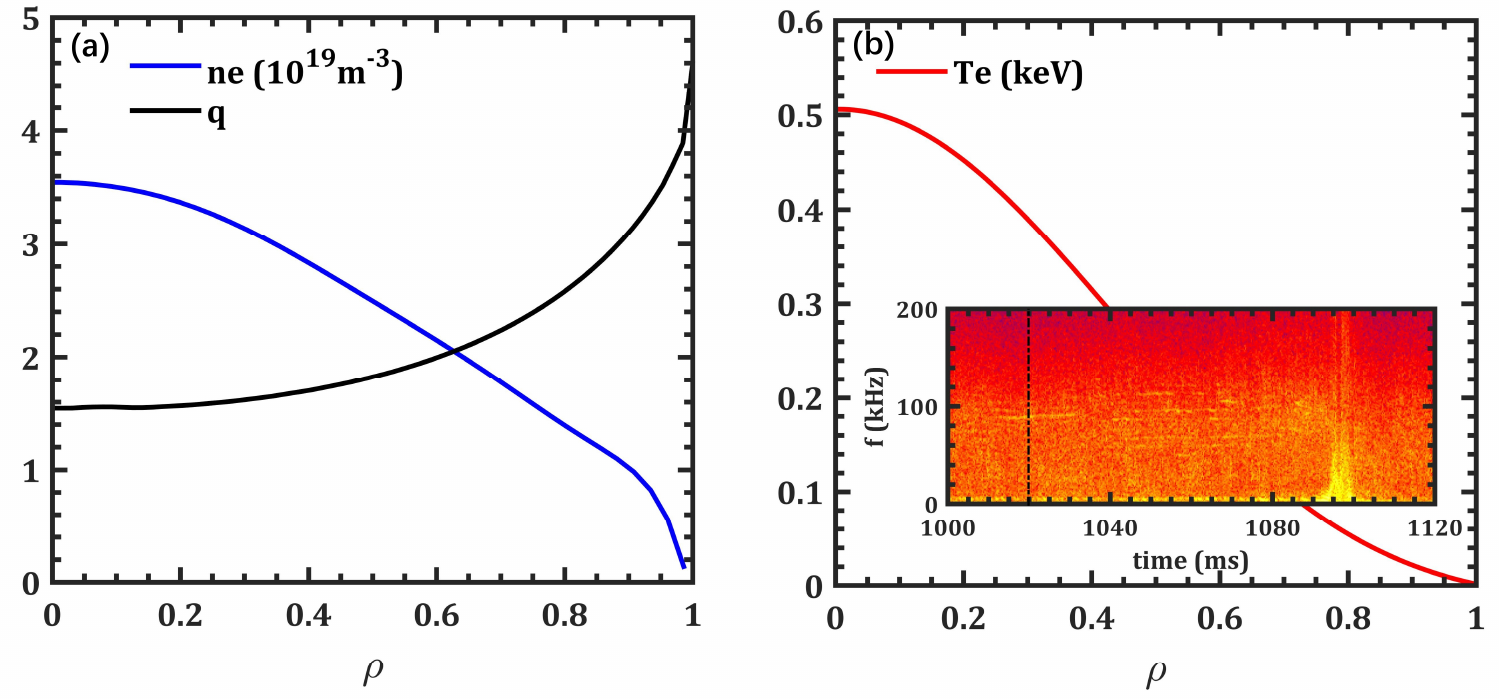}
 \caption{(a) Electron density $ne$ profile and safety factor $q$ profile for shot \#38524 at 1020 ms. (b) Electron temperature $Te$ profile provided by ECE and spectrum of microwave interferometry signal. The black dashed line indicates the time (1020 ms) at which the profiles are shown. The instability frequency at this time is approximately 90 kHz.}
\label{fig10}
\end{figure}

 \begin{figure}[!htbp]
 \centering
        \includegraphics[width=0.9\linewidth]{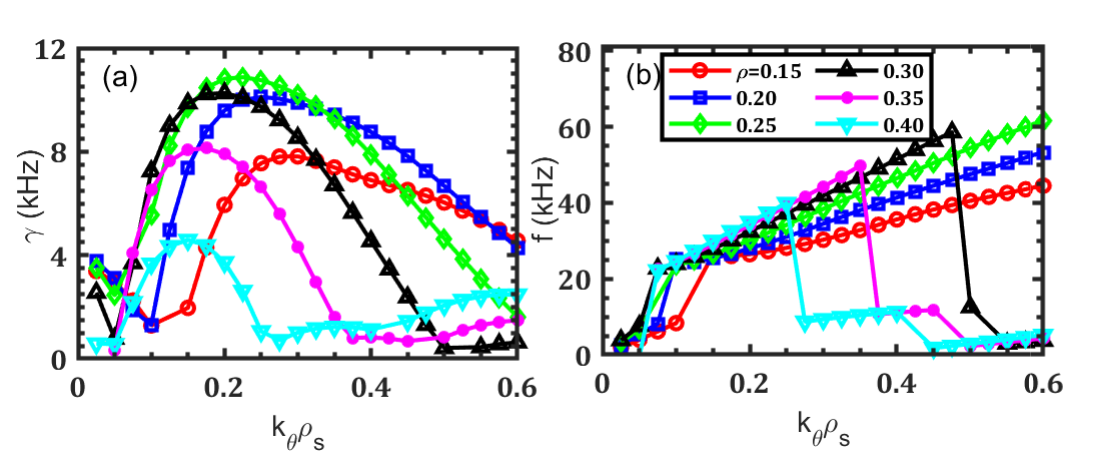}
 \caption{Linear GENE simulation results for the KBM instability using experimental profiles from shot \#38524 at 1020 ms. (a) Growth rate $\gamma$ (b) real frequency $f$are plotted against the normalized poloidal wavenumber $k_\theta \rho_s$, where $\rho_s$ is the ion sound gyrororadius. Each color represents a different normalized radius $\rho=r/a$.}
\label{fig11}
\end{figure}

To identify the core instabilities in high-density plasmas, numerical simulations were carried out using experimental parameters from a representative discharge. Figure \ref{fig10} shows the radial profiles of electron density $ne$ and electron temperature $Te$ and safety factor $q$ at 1020 ms for shot \#38524. Given that it is an ohmic heating plasma and ion temperature data is lacking, it is assumed in the simulation that the ion temperature equals the electron temperature. Through estimation, at the moment depicted in Figure \ref{fig10}, the energy exchange time between electrons and ions in the plasma, $\tau_{ei}$, is on the order of milliseconds, which is much shorter than the energy confinement time $\tau_E$. Therefore, it can be assumed that the ion temperature is roughly equal to the electron temperature. The spectrum of the microwave interferometry signal for shot \#38524 is also depicted in Figure \ref{fig10}(b). It is noteworthy that in Figure \ref{fig10}(a), the black solid line represents the safety factor profile. As evident, the safety factor profile for shot \#38524 precisely matches the previously described safety factor profile. At 1020 ms, an instability is observed near 90 kHz.
These profile data were used as input to the GENE code \cite{Jenko_2000} to compute linear stability. Figure \ref{fig11} shows the result of growth rate $\gamma$ and real frequency $f$ as functions of normalized poloidal wavenumber $k_\theta \rho_s$. And the different color lines in Figure \ref{fig11} indicate different normalized radius $\rho$. Under the parameter conditions shown in Figure \ref{fig10}, it is estimated that the ion diamagnetic drift frequency $f_{*pi}$ and the ion transit frequency $f_{ti}$ are approximately 20 kHz. Therefore, it can be concluded from the frequency characteristics that the dominant modes in Figure \ref{fig11} are the KBM/AITGs. The simulations show that KBM/AITG is unstable under the conditions of shot \#38524 at 1020 ms. As seen in Figure \ref{fig11}(a), KBM/AITG modes are unstable over a broad radial range, e.g., 0.15 $< \rho <$ 0.4. The peak growth rate occurs at $k_\theta\rho_s\sim 0.25$ for $\rho=0.25$, with a corresponding real frequency of about 32 kHz. The frequency difference between the instabilities in Figure \ref{fig10} and Figure \ref{fig11} can be attributed to Doppler shift. Due to the lack of effective diagnostic data for the toroidal rotation frequency in Ohmic-heated plasma, a direct comparison between the experimental and simulated mode frequencies could not be established. The eigenmode structure obtained from GENE simulations exhibits only slight frequency variation across radius, consistent with the spectral behavior observed in Figure \ref{fig10}.

 \begin{figure}[!t]
 \centering
        \includegraphics[scale=0.8]{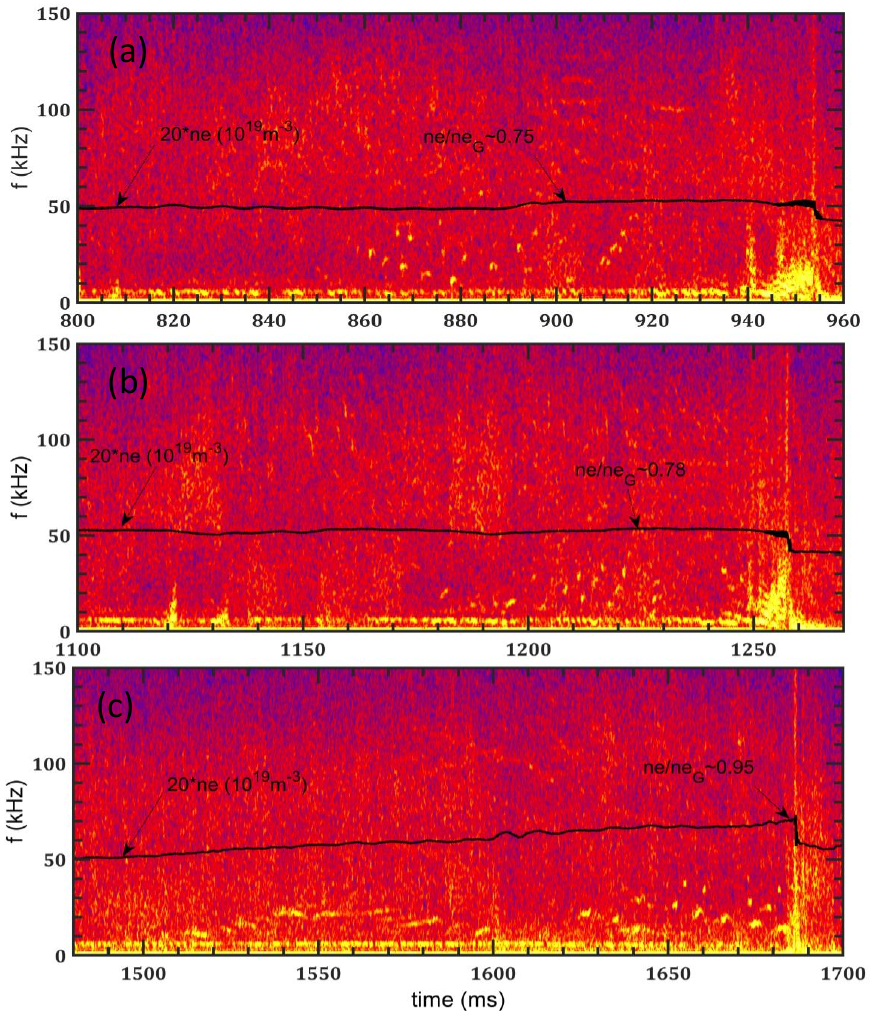}
 \caption{Microwave interferometry spectra are shown for three discharges: (a) \#38635, (b) \#38638, and (c) \#38273.  The black line in each panel represents 20 times of electron density $20*ne$ and the unit is $10^{19} \mathrm{m^{-3}}$. Black arrows point to the times when "Christmas-light" modes emerge in the spectrum.}
\label{fig12}
\end{figure}

 \begin{figure}[!htbp]
 \centering
        \includegraphics[scale=0.07]{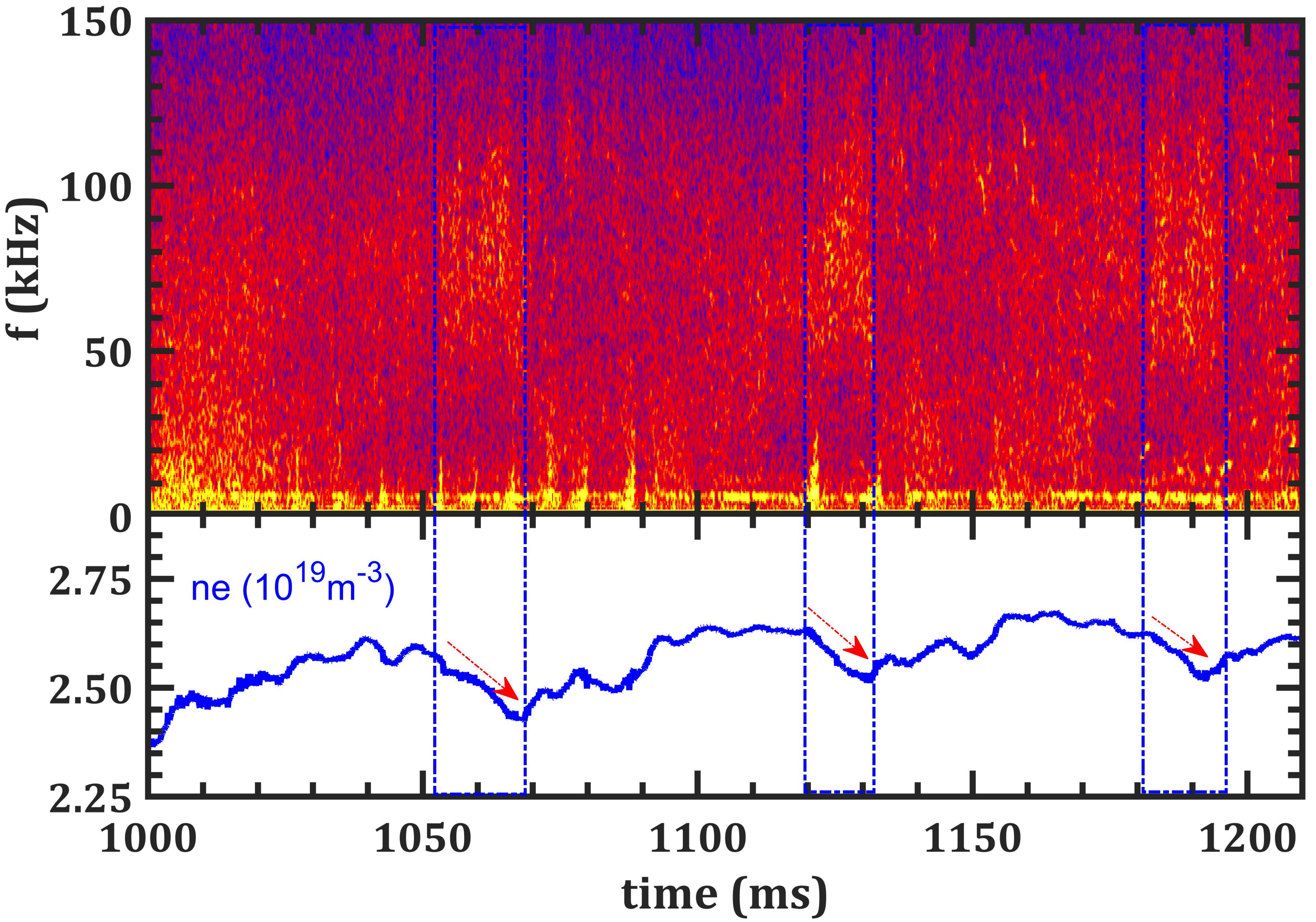}
 \caption{Microwave interferometry spectrum and the corresponding line-averaged density trace (blue) for shot \#38638. Blue dashed lines mark the time interval during which core-localized modes appear. Red arrows indicate the trend in line-averaged density following the onset of these modes.}
\label{fig13}
\end{figure}

Through GENE simulations, the clMHD modes observed in high-density Ohmic-heated HL-2A plasmas have been identified as KBM/AITG-type modes. Next, we will discuss the impact of KBM/AITG modes on the plasma. Figure \ref{fig12} presents additional examples of high-density ohmic discharges in which only low-frequency instabilities ( < 50 kHz) appear in the microwave interferometry spectra. As shown in panels (a)–(c), these low-frequency modes emerge once the Greenwald fraction $ne/ne_G$ reaches certain thresholds—ranging from 0.75 to 0.95—and generally precede density-limit disruptions. These modes closely resemble the "Christmas-light" modes reported on DIII-D \cite{Heidbrink_2021}.

 The frequency variations among instabilities result from differences in the toroidal mode number $n$. As evidenced by DIII-D results, the mode frequencies in the plasma frame remain nearly constant. According to the relation $f_{lab}=f_{plasma} + nf_{v\phi}$, the frequency shifts seen in Figure \ref{fig12} are due to a decrease in $n$. A key question is what causes this decrease in toroidal mode number. In DIII-D experiments, such variations are linked to changes in the minimum safety factor $q_{min}$. Based on the experimental data from DIII-D, the variation in the toroidal mode number $n$ originates from changes in the minimum safety factor $q_{min}$. However, in the high-density ohmic heating plasmas studied here, reversed magnetic shear seldom occurs, so $q_{min}$ remains at the plasma center. This results in a weak magnetic shear configuration in the core region, as illustrated in Fig. \ref{fig10}(a). Therefore, the change in $n$ must originate from local variations in the safety factor profile $q(r)$ at the mode location.
 
  The difference in toroidal mode number between two adjacent modes is 1, so the frequency spacing between them equals the toroidal rotation frequency. Taking Fig. \ref{fig12}(c) as an example, modes appear as bright spots in the spectrum. Those that align along a straight line are considered a group; the frequency spacing between adjacent modes in the same group is about 3.3 kHz, implying a local toroidal rotation frequency of the same value.  The last mode in this group appears at approximately 15.6 kHz, and within 5 ms after its emergence, a minor density-limit disruption occurs. Since the plasma-frame frequency $f_{plasma}>0$ , the toroidal mode number for this mode must satisfy $n<5$.

 The radial width of an Alfvénic instability in a region of weak magnetic shear can be estimated as $\Delta_m\sim r_m/(n^{1/2}Sq_m)$\cite{Zonca_Briguglio_2002}, where $r_m$ is the minor radius at the location of the mode, $n$ is the toroidal mode number of the mode, $S=\sqrt{r_m^2q_m^{''}/q_m^2}$ is the generalized shear value, and $q_m$ is the safety factor at the location of the mode. Assuming that $r_m, q_m$ and $S$ do not change significantly over short timescales, a decrease in $n$ leads to an increase in the mode’s radial width. This broadening appears linked to the onset of density-limit disruptions, as shown in Figure \ref{fig12}(a)–(c): disruptions tend to occur after $n$ has decreased sufficiently—that is, after the modes have radially expanded. A plausible explanation is that these clMHD instabilities degrade particle or energy confinement, and their effect strengthens as their radial width increases, eventually triggering a disruption. Their role in particle transport will be examined in detail later.

The high-frequency instabilities ($\sim$100 kHz) shown in Figure \ref{fig5}(a) exhibit dynamics similar to those of low-frequency KBM/AITG modes. Their frequencies satisfy $f\sim f_{TAE} >f_{ti}$, where $f_{ti}=v_\parallel/(2\pi qR)$ represents the transit frequency of passing ions. It is expected that finite compressibility of the core plasma will significantly influence mode dynamics via resonant interaction with ion transit motion along magnetic field lines \cite{Zonca_Chen_Santoro_Dong_1998}. Since the fraction of trapped particles in a tokamak scales as $f_{trap}=(\frac{2r}{R_0+r})^{1/2}$, the majority of particles in the core region are passing particles. The transit frequency of thermal ions resonates with KBM dynamics, leading to $f_{KBM}\sim f_{ti}$. High-frequency instabilities in the core may similarly interact with thermal passing ions. The difference in frequencies between high- and low-frequency modes may result from different toroidal mode numbers and different mode location, producing discrete features in the frequency spectrum.

The experimental data presented in Figure \ref{fig13} provide evidence of the role played by KBM/AITG modes in electron transport on the HL-2A tokamak. The figure displays the microwave interferometry spectrum alongside the line-averaged density for shot \#38638, with blue dashed lines highlighting intervals of KBM/AITG activity. A clear reduction in plasma density is observed following the emergence of these modes, which is likely due to enhanced particle transport in the core region induced by the KBM/AITGs. Consequently, these instabilities are implicated in limiting the achievable plasma beta, as they lead to a reduction in both electron temperature and density.

\section{Summary}
In Ohmic-heated plasmas in the HL-2A tokamak, the Greenwald density limit was exceeded by means of gas puffing, achieving a maximum density fraction of $ne/ne_G\sim1.5$ for a duration of nearly 500 ms, or about $30\tau_E$. Some clMHD modes are observed prior to both major and minor disruptions and are considered to contribute to their onset. Comparative analysis of soft X-ray signals, magnetic probe signals, and density profiles indicated that these clMHD modes are easiler to be unstable when central safety factor is larger than 1. The clMHD modes identified in spectra are found to limit the achievable plasma density. In microwave interferometry spectra, broadband perturbations are also observed as plasma density rises. Spectral broadening increases with rising plasma density, with the maximum perturbation frequency reaching about 200 kHz before a density-limit disruption occurs. 

The clMHD modes have been identified as KBM/AITG type through simulations with the GENE code. Statistical analysis of experimental data further shows that their excitation correlates with a threshold in the Greenwald density fraction, approximately $ne/ne_G\sim0.7$ in HL-2A Ohmic-heated plasmas. It is also observed that the toroidal mode number of these KBM/AITGs decreases with increasing density, corresponding to a broadening of the modes' radial width. Experimentally, KBMs/AITGs have been shown to directly reduce the line-averaged plasma density, indicating that they can enhance particle transport. Consequently, KBM/AITGs play a critical role in limiting the plasma density in high-density Ohmic discharges. These results provide important insight into the mechanisms underlying the density limit.


%
%

\ack{One of the authors (W. Chen) is very grateful to the HL-2A group. This work is supported in part by NSFC under Grants No. 12125502, 12375211, 12475215 and 12205034, by the National MCF Energy R\&D Program under Grants No. 2024YFE03190004, by SWIP innovation under Grants No. 202301XWCX001, and by the Nature Science Foundation of Sichuan Province under Grants No.2025ZNSFSC0833 and 2025ZNSFSC0059. }

%
%
%

\bibliographystyle{unsrt}    
\bibliography{Bibref}           

@article{Greenwald_1988, title={A new look at density limits in tokamaks}, volume={28}, ISSN={0029-5515, 1741-4326}, DOI={10.1088/0029-5515/28/12/009},  number={12}, journal={Nuclear Fusion}, author={Greenwald. M. et al.}, year={1988}, pages={2199–2207}, language={en} }

@article{Greenwald_2002, title={Density limits in toroidal plasmas}, volume={44}, ISSN={07413335}, DOI={10.1088/0741-3335/44/8/201},  number={8}, journal={Plasma Physics and Controlled Fusion}, author={Greenwald. M.}, year={2002}, pages={R27–R53}, language={en} }

@article{Kirneva_2015, title={High density experiments in {TCV} ohmically heated and {L}-mode plasmas}, volume={57}, ISSN={0741-3335, 1361-6587}, DOI={10.1088/0741-3335/57/2/025002},  journal={Plasma Physics and Controlled Fusion},author={Kirneva. N.A. et al.}, year={2015}, pages={025002}, language={en} }

@article{Pucella_2013, title={Density limit experiments on {FTU}}, volume={53}, ISSN={0029-5515, 1741-4326}, DOI={10.1088/0029-5515/53/8/083002}, abstractNote={One of the main problems in tokamak fusion devices concerns the capability to operate at a high plasma density, which is observed to be limited by the appearance of catastrophic events causing loss of plasma conﬁnement. The commonly used empirical scaling law for the density limit is the Greenwald limit, predicting that the maximum achievable lineaveraged density along a central chord depends only on the average plasma current density. However, the Greenwald density limit has been exceeded in tokamak experiments in the case of peaked density proﬁles, indicating that the edge density is the real parameter responsible for the density limit. Recently, it has been shown on the Frascati Tokamak Upgrade (FTU) that the Greenwald density limit is exceeded in gas-fuelled discharges with a high value of the edge safety factor. In order to understand this behaviour, dedicated density limit experiments were performed on FTU, in which the high density domain was explored in a wide range of values of plasma current (Ip = 500–900 kA) and toroidal magnetic ﬁeld (BT = 4–8 T). These experiments conﬁrm the edge nature of the density limit, as a Greenwald-like scaling holds for the maximum achievable line-averaged density along a peripheral chord passing at r/a 4/5. On the other hand, the maximum achievable line-averaged density along a central chord does not depend on the average plasma current density and essentially depends on the toroidal magnetic ﬁeld only. This behaviour is explained in terms of density proﬁle peaking in the high density domain, with a peaking factor at the disruption depending on the edge safety factor. The possibility that the MARFE (multifaced asymmetric radiation from the edge) phenomenon is the cause of the peaking has been considered, with the MARFE believed to form a channel for the penetration of the neutral particles into deeper layers of the plasma. Finally, the magnetohydrodynamic (MHD) analysis has shown that also the central line-averaged density at the onset of the MHD activity depends only on the toroidal magnetic ﬁeld.}, number={8}, journal={Nuclear Fusion}, author={Pucella. G. et al.}, year={2013},  pages={083002}, language={en} }

@article{White_Gates_Brennan_2015, title={Thermal island destabilization and the {G}reenwald limit}, volume={22}, ISSN={1070-664X, 1089-7674}, DOI={10.1063/1.4913433}, abstractNote={Magnetic reconnection is ubiquitous in the magnetosphere, the solar corona, and in toroidal fusion research discharges. In a fusion device, a magnetic island saturates at a width which produces a minimum in the magnetic energy of the configuration. At saturation, the modified current density profile, a function of the flux in the island, is essentially flat, the growth rate proportional to the difference in the current at the O-point and the X-point. Further modification of the current density profile in the island interior causes a change in the island stability and additional growth or contraction of the saturated island. Because field lines in an island are isolated from the outside plasma, an island can heat or cool preferentially depending on the balance of Ohmic heating and radiation loss in the interior, changing the resistivity and hence the current in the island. A simple model of island destabilization due to radiation cooling of the island is constructed, and the effect of modification of the current within an island is calculated. An additional destabilization effect is described, and it is shown that a small imbalance of heating can lead to exponential growth of the island. A destabilized magnetic island near the plasma edge can lead to plasma loss, and because the radiation is proportional to plasma density and charge, this effect can cause an impurity dependent density limit.}, number={2}, journal={Physics of Plasmas}, author={White. R. B. et al.}, year={2015}, pages={022514}, language={en} }

@article{Teng_Brennan_2016, title={A predictive model for the tokamak density limit}, volume={56}, ISSN={0029-5515, 1741-4326}, DOI={2023020716565200}, abstractNote={The Greenwald density limit, found in all tokamak experiments, is reproduced for the first time using a phenomenologically correct model with parameters in the range of experiments. A simple model of equilibrium evolution and local power balance inside the island has been implemented to calculate the radiation-driven thermo-resistive tearing mode growth and explain the density limit. Strong destabilization of the tearing mode due to an imbalance of local Ohmic heating and radiative cooling in the island predicts the density limit within a few percent. The density limit is found to be a local edge limit and weakly dependent on impurity densities. Results are robust to a substantial variation in model parameters within the range of experiments.}, number={10}, journal={Nuclear Fusion}, author={Teng. Q. et al.}, year={2016}, pages={106001}, language={en} }

@article{Giacomin_Ricci_2022, title={Turbulent transport regimes in the tokamak boundary and operational limits}, volume={29}, ISSN={1070-664X, 1089-7674}, DOI={10.1063/5.0090541}, abstractNote={Two-fluid, three-dimensional, flux-driven, global, electromagnetic turbulence simulations carried out by using the GBS code are used to identify the main parameters controlling turbulent transport in the tokamak boundary and to delineate an electromagnetic phase space of edge turbulence. Four turbulent transport regimes are identified: (i) a regime of fully developed turbulence appearing at intermediate values of collisionality and $beta$, with turbulence driven by resistive ballooning modes, related to the L-mode operation of tokamaks, (ii) a regime of reduced turbulent transport at low collisionality and large heat source, with turbulence driven by drift-waves, related to a high-density H-mode regime, (iii) a regime of extremely large turbulent transport at high collisionality, which is associated with the crossing of the density limit, and (iv) a regime above the ideal ballooning limit at high $beta$, with global modes affecting the dynamics of the entire confined region, which can be associated with the crossing of the $beta$ limit. The transition from the reduced to the developed turbulent transport regime is associated here with the H-mode density limit and an analytical scaling law for maximum edge density achievable in H-mode is obtained. Analogously, analytical scaling laws for the crossing of the L-mode density and $beta$ limits are provided and compared to the results of GBS simulations.}, number={6}, journal={Physics of Plasmas}, author={Giacomin. M. et al.}, year={2022}, pages={062303}, language={en} }

@article{Giacomin_Ricci_2020, title={Investigation of turbulent transport regimes in the tokamak edge by using two-fluid simulations}, volume={86}, ISSN={0022-3778, 1469-7807}, DOI={10.1017/S0022377820000914}, abstractNote={The results of ﬂux-driven, two-ﬂuid simulations in single-null conﬁgurations are used to investigate the processes determining the turbulent transport in the tokamak edge. Three turbulent transport regimes are identiﬁed: (i) a developed transport regime with turbulence driven by an interchange instability, which shares a number of features with the standard L-mode of tokamak operation; (ii) a suppressed transport regime, characterized by a higher value of the energy conﬁnement time, low-amplitude relative ﬂuctuations driven by a Kelvin–Helmholtz instability, a strong E × B sheared ﬂow and the formation of a transport barrier, which recalls the H-mode; and (iii) a degraded conﬁnement regime, characterized by a catastrophically large interchange-driven turbulent transport, which recalls the crossing of the Greenwald density limit. We derive an analytical expression of the pressure gradient length in the three regimes. The transition from the developed transport regime to the suppressed transport regime is obtained by increasing the heat source or decreasing the collisionality and vice versa for the transition from the developed transport regime to the degraded conﬁnement regime. An analytical expression of the power threshold to access the suppressed transport regime, linked to the power threshold for H-mode access, as well as the maximum density achievable before entering the degraded conﬁnement regime, related to the Greenwald density, are also derived. The experimental dependencies of the power threshold for H-mode access on density, tokamak major radius and isotope mass are retrieved. The analytical estimate of the density limit contains the correct dependence on the plasma current and on the tokamak minor radius.}, number={5}, journal={Journal of Plasma Physics}, author={Giacomin. M. et al.}, year={2020}, pages={905860502}, language={en} }

@article{Rapp_1999, title={Density limits in {TEXTOR}-94 auxiliary heated discharges}, volume={39}, ISSN={0029-5515}, DOI={10.1088/0029-5515/39/6/305}, abstractNote={In a tokamak plasma the maximum achievable density is limited. A too high density will result in a violent end of a discharge. Two types of density limit disruption can be distinguished: (a) impure and moderately heated discharges, if the radiative power exceeds the input power, (b) clean, auxiliary heated discharges, where the Greenwald limit is encountered. It has been found that in TEXTOR-94 these two density limits diﬀer by the radiative instability in the plasma boundary, which preceeds the disruption. A symmetric radiative mantle and a detachment are observed prior to the ﬁrst type, while the Greenwald limit has a MARFE precursor. Control of the impurity content, edge and recycling properties prevents the growth of the MARFE and makes it possible to exceed the Greenwald limit in TEXTOR-94 by more than a factor of 2. High densities have been obtained by means of normal gas feed. Maximum central densities of ne(0) = 1.3 × 1020 m−3 have been obtained. The maximum achievable density scales with the input power and plasma current. Non-disruptive discharges, with a stationary (t > 25τE) density a factor of 1.93 above the Greenwald limit have been produced in L mode. The radiative losses and impurity concentration have been maintained at a relatively low level during the entire high density phase.}, number={6}, journal={Nuclear Fusion}, author={Rapp. J. et al.}, year={1999},  pages={765–776}, language={en} }

@article{Lang_Suttrop_2012, title={High-density {H}-mode operation by pellet injection and {ELM} mitigation with the new active in-vessel saddle coils in {ASDEX U}pgrade}, volume={52}, ISSN={0029-5515, 1741-4326}, DOI={10.1088/0029-5515/52/2/023017}, abstractNote={Recent experiments at ASDEX Upgrade demonstrate the compatibility of ELM mitigation by magnetic perturbations with efficient particle fuelling by inboard pellet injection. ELM mitigation persists in a high density, high collisionality regime even with the strongest applied pellet perturbations. Pellets injected into mitigation phases trigger no type-I ELM like events unlike when launched into unmitigated type-I ELMy plasmas. Furthermore, the absence of ELMs results in improved fuelling efficiency and persistent density build up. Pellet injection is helpful to access the ELM-mitigation regime by raising the edge density beyond the required threshold level, mostly eliminating the need for strong gas puff. Finally, strong pellet fuelling can be applied to access high densities beyond the density limit encountered with pure gas puffing. Core densities of up to 1.6 times the Greenwald density have been reached while maintaining ELM mitigation. No upper density limit for the ELM-mitigated regime has been encountered so far; limitations were set solely by technical restrictions of the pellet launcher. Reliable and reproducible operation at line averaged densities from 0.75 up to 1.5 times the Greenwald density is demonstrated using pellets. However, in this density range there is no indication of the positive confinement dependence on density implied by the ITERH98P(y,2) scaling.}, number={2}, journal={Nuclear Fusion}, author={Lang. P.T. et al.}, year={2012}, pages={023017}, language={en} }

@article{Bell_1992, title={Attainment of high plasma densities in {TFTR} with injection of multiple deuterium pellets}, volume={32}, ISSN={0029-5515}, DOI={10.1088/0029-5515/32/9/I06}, abstractNote={Sequential injection of multiple deuterium pellets combined with high power neutral beam heating in TFTR has produced stable plasmas with very high central densities, up to 5 X lo2’ m-3. The line averaged densities achieved in these experiments correspond to Murakami parameters ii,R/B, up to 12 X l O I 9 m-*.T-’. This value, which does not appear to represent an intrinsic density limit, surpasses the values previously obtained in TFTR and exceeds, by more than a factor of two, the density limit predicted by the scaling of Greenwald et al. (Nuclear Fusion 28 (1988) 2199). The highest densities obtained with both pellet and gas fuelling have been achieved since boronization was applied to the first wall in TFTR. The characteristics and energy balance of the highest density plasma are discussed.}, number={9}, journal={Nuclear Fusion}, author={Bell. M. G. et al.}, year={1992}, pages={1585–1591}, language={en} }

@article{duan2009overview,
  title={Overview of experimental results on {HL-2A}},
  author={Duan. X. R. et al.},
  journal={Nuclear fusion},
  volume={49},
  number={10},
  pages={104012},
  year={2009},
  publisher={IOP Publishing}
}

@article{li2017new,
  title={A new high sensitivity far-infrared laser interferometer for the {HL-2A} tokamak},
  author={Li. Y. G. et al.},
  journal={Review of Scientific Instruments},
  volume={88},
  number={8},
  year={2017},
  publisher={AIP Publishing}
}

@article{shi2014calibration,
  title={Calibration of a 32 channel electron cyclotron emission radiometer on the {HL-2A} tokamak},
  author={Shi. Z. B. et al.},
  journal={Review of Scientific Instruments},
  volume={85},
  number={2},
  year={2014},
  publisher={AIP Publishing}
}

@article{Shi_2016, title={Multichannel Microwave Interferometer for Simultaneous Measurement of Electron Density and its Fluctuation on {HL-2A} Tokamak}, volume={18}, ISSN={1009-0630}, DOI={10.1088/1009-0630/18/7/02}, abstractNote={A multichannel microwave interferometer system has been developed on the HL2A tokomak. Its working frequency is well designed to avoid the fringe jump eﬀect. Taking the structure of HL-2A into account, its antennas are installed in the horizontal direction, i.e. one launcher in high ﬁeld side (HFS) and four receivers in low ﬁeld side (LFS). The fan-shaped measurement area covers those regions where the magnetohydrodynamics (MHD) instabilities are active. The heterodyne technique contributes to its high temporal resolution (1 µs). It is possible for the multichannel system to realize simultaneous measurements of density and its ﬂuctuation. The quadrature phase detection based on the zero-crossing method is introduced to density measurement. With this system, reliable line-averaged densities and density proﬁles are obtained. The location of the saturated internal kink mode can be ﬁgured out from the mode showing diﬀerent intensities on four channels, and the result agrees well with that measured by electron cyclotron emission imaging (ECEI).}, number={7}, journal={Plasma Science and Technology}, author={Shi. P. W. et al.}, year={2016}, pages={708–713}, language={en} }

@article{Nazikian_1998, title={Toroidal Alfvén eigenmodes in {TFTR} deuterium–tritium plasmas}, volume={5}, ISSN={1070-664X, 1089-7674}, DOI={10.1063/1.872818}, abstractNote={Purely alpha-particle-driven toroidal Alfvén eigenmodes (TAEs) with toroidal mode numbers n=1–6 have been observed in deuterium–tritium (D–T) plasmas on the tokamak fusion test reactor [D. J. Grove and D. M. Meade, Nucl. Fusion 25, 1167 (1985)]. The appearance of mode activity following termination of neutral beam injection in plasmas with q(0)&gt;1 is generally consistent with theoretical predictions of TAE stability [G. Y. Fu et al. Phys. Plasmas 3, 4036 (1996)]. Internal reflectometer measurements of TAE activity is compared with theoretical calculations of the radial mode structure. Core localization of the modes to the region of reduced central magnetic shear is confirmed, however the mode structure can deviate significantly from theoretical estimates. The peak measured TAE amplitude of δn/n∼10−4 at r/a∼0.3−0.4 corresponds to δB/B∼10−5, while δB/B∼10−8 is measured at the plasma edge. Enhanced alpha particle loss associated with TAE activity has not been observed.}, number={5}, journal={Physics of Plasmas}, author={Nazikian. R. et al.}, year={1998},  pages={1703–1711}, language={en} }

@article{strait1994doppler,
  title={Doppler shift of the TAE mode frequency in {DIII-D}},
  author={Strait. E. J. et al.},
  journal={Plasma physics and controlled fusion},
  volume={36},
  number={7},
  pages={1211},
  year={1994},
  publisher={IOP Publishing}
}

@book{freidberg2008plasma,
  title={Plasma physics and fusion energy},
  author={Freidberg. J. P},
  year={2008},
  pages={49--56},
  publisher={Cambridge university press}
}

@article{Chen_2016, title={Core-localized {A}lfvénic modes driven by energetic ions in {HL-2A NBI} plasmas with weak magnetic shears}, volume={56}, ISSN={0029-5515, 1741-4326}, DOI={10.1088/0029-5515/56/3/036018}, abstractNote={Recent experimental results that are associated with the core-localized (i.e. normalized radius ρ = r /a < 0.5) Alfvénic modes in HL-2A neutral beam injection (NBI) plasmas with weak magnetic shears are reported. In the different plasma parameter regions, the energetic ions produced by the NBI drive multiple Alfvénic instabilities, such as the toroidal Alfvén eigenmode (TAE), beta-induced Alfvén eigenmode, reversed shear Alfvén eigenmode (RSAE) and fishbone and energetic particle mode (EPM). Here, we focus on the high-frequency RSAE (HFRSAE) and resonant kinetic ballooning mode (rKBM). A group of downwardsweeping frequency coherent modes (HFRSAEs) with 100  <  f  <  500 kHz and n  =  3–7 are often observed with an increase in the edge safety factor, qa. Their measured frequency is more than that of the TAEs, and fmin ∼ fTAE. The analysis suggests that these modes localize inside the high-order Alfvén eigenmode (AE) gap of the Alfvénic continuum, and their eigenfrequency and eigenfunction depend on the qmin and q-profile. When the core plasma density is more than ne0 > 3.0 × 1019 m−3 and the impurity or supersonic molecular beam enters the bulk plasma, the profiles of the plasma density/pressure peak, and the magnetic shear is weak or negative. In this case, a group of multi-harmonic coherent modes (rKBMs) with 30  <  f  <  150 kHz and n  =  2–9 are observed through multiple diagnostic techniques, and f∗pi /2 < fMHD = flab − nfvφ < f∗pi, where f∗pi = ω∗pi/2π is the diamagnetic drift frequency of the thermal ion. It is found that the HFRSAEs can transit into the rKBMs when the density profile suddenly peaks. Neutron monitoring outside the vacuum chamber demonstrates that the HFRSAE and rKBM both degrade the confinement of the energetic ions. The rKBM instabilities also affect the bulk plasma performance.}, number={3}, journal={Nuclear Fusion}, author={Chen. W. et al.}, year={2016}, pages={036018}, language={en} }

@article{Chen_2018, title={Kinetic electromagnetic instabilities in an {ITB} plasma with weak magnetic shear}, volume={58}, ISSN={0029-5515, 1741-4326}, DOI={10.1088/1741-4326/aaaece}, abstractNote={Kinetic Alfvén and pressure gradient driven instabilities are very common in magnetized plasmas, both in space and the laboratory. These instabilities will be easily excited by energetic particles (EPs) and/or pressure gradients in present-day fusion and future burning plasmas. This will not only cause the loss and redistribution of the EPs, but also affect plasma confinement and transport. Alfvénic ion temperature gradient (AITG) instabilities with the frequency ωBAE < ω < ωTAE and the toroidal mode numbers n = 2−8 are found to be unstable in NBI internal transport barrier plasmas with weak shear and low pressure gradients, where ωBAE and ωTAE are the frequencies of the beta- and toroidicity-induced Alfvén eigenmodes, respectively. The measured results are consistent with the general fishbone-like dispersion relation and kinetic ballooning mode equation, and the modes become more unstable the smaller the magnetic shear is in low pressure gradient regions. The interaction between AITG activity and EPs also needs to be investigated with greater attention in fusion plasmas, such as ITER (Tomabechi and The ITER Team 1991 Nucl. Fusion 31 1135), since these fluctuations can be enhanced by weak magnetic shear and EPs.}, number={5}, journal={Nuclear Fusion}, author={Chen. W. et al.}, year={2018}, pages={056004}, language={en} }

@article{ma2023low,
  title={Low-frequency shear {A}lfv{\'e}n waves at {DIII-D}: Theoretical interpretation of experimental observations},
  author={Ma. R. R. et al.},
  journal={Physics of Plasmas},
  volume={30},
  number={4},
  year={2023},
  publisher={AIP Publishing}
}

@article{Zonca_Chen_Santoro_1996, title={Kinetic theory of low-frequency {A}lfvén modes in tokamaks}, volume={38}, ISSN={0741-3335, 1361-6587}, DOI={10.1088/0741-3335/38/11/011}, abstractNote={The kinetic theory of low-frequency Alfve´n modes in tokamaks is presented. The inclusion of both diamagnetic effects and ﬁnite core-plasma ion compressibility generalizes previous theoretical analyses (Tsai S T and Chen L 1993 Phys. Fluids B 5 3284) of kinetic ballooning modes and clariﬁes their strong connection to beta-induced Alfve´n eigenmodes. The derivation of an analytic mode dispersion relation allows us to study the linear stability of both types of modes as a function of the parameters characterizing the local plasma equilibrium and to demonstrate that the most unstable regime corresponds to a strong coupling between the two branches due to the ﬁnite thermal ion temperature gradient. In addition, we also show that, under certain circumstances, non-collective modes may be present in the plasma, formed as a superposition of local oscillations which are quasi-exponentially growing in time.}, number={11}, journal={Plasma Physics and Controlled Fusion}, author={Zonca. F. et al.}, year={1996}, pages={2011–2028}, language={en} }

@article{Hou_2018, title={{NIMROD} calculations of energetic particle driven toroidal {A}lfvén eigenmodes}, volume={25}, ISSN={1070-664X, 1089-7674}, DOI={10.1063/1.4999619}, abstractNote={Toroidal Alfvén eigenmodes (TAEs) are gap modes induced by the toroidicity of tokamak plasmas in the absence of continuum damping. They can be excited by energetic particles (EPs) when the EP drive exceeds other dampings, such as electron and ion Landau damping, and collisional and radiative damping. A TAE benchmark case, which was proposed by the International Tokamak Physics Activity group, is studied in this work. The numerical calculations of linear growth of TAEs driven by EPs in a circular-shaped, large aspect ratio tokamak have been performed using the Hybrid Kinetic-MHD (HK-MHD) model implemented in the NIMROD code. This HK-MHD model couples a δf particle-in-cell representation of EPs with the 3D MHD representation of the bulk plasma through moment closure for the momentum conservation equation. Both the excitation of TAEs and their transition to energetic particle modes (EPMs) have been observed. The influence of EP density, temperature, density gradient, and position of the maximum relative density gradient, on the frequency and the growth rate of TAEs are obtained, which are consistent with those from the eigen-analysis calculations, kinetic-MHD, and gyrokinetic simulations for an initial Maxwellian distribution of EPs. The relative pressure gradient of EP at the radial location of the TAE gap, which represents the drive strength of EPs, can strongly affect the growth rate of TAEs. It is demonstrated that the mode transition due to EP drive variation leads to not only the change of frequency but also the change of the mode structure. This mechanism can be helpful in understanding the nonlinear physics of TAE/EPM, such as frequency chirping.}, number={1}, journal={Physics of Plasmas}, author={Hou. Y. W. et al.}, year={2018}, pages={012501}, language={en} }

@article{Stabler_1992, title={Density limit investigations on {ASDEX}}, volume={32}, ISSN={0029-5515}, DOI={10.1088/0029-5515/32/9/I05}, abstractNote={Density limit investigations on ASDEX have been performed under a variety of conditions: ohmically heated and neutral injection heated plasmas in H,, D, and He have been studied in different divertor configurations, after various wall coating procedures, with gas puff and pellet fuelling, and in different confinement regimes with their characteristically different density profiles. A detailed description of the parametric dependence of the density limit, which in all cases is a disruptive limit, is given. This limit is shown to be a limit to the density at the plasma edge. Therefore, the highest densities corresponding to &Rq,/B, > 30 x l O I 9 m-’.T-’ are obtained with centrally peaked ne profiles. Radiation from the main plasma at the density limit is always significantly below the total input power. The plasma disruption is due to an m = 2 instability which for medium and high q, is preceded by one or more minor disruptions. In this range of q,, the disruptive instability is initiated by the occurrence of a Marfe on the high field side as a consequence of strong plasma cooling in this region. The duration of the Marfe increases with increasing distance between the plasma edge and the q = 2 surface. After penetrating onto closed flux surfaces the Marfe leads to a current contraction and a subsequent destabilization of the m = 2 mode. In helium plasmas a strongly radiating, poloidally symmetric shell is observed before the density limit instead of a Marfe. An instantaneous destabilization of this mode is observed at low 9,. Detailed measurements of plasma edge and divertor parameters close to the density limit indicate the development of a cold, dense divertor plasma before the disruption. Models describing the scrape-off layer and,the divertor region predict an upper limit to the edge density at low divertor temperatures according to power balance considerations. Their relations to the experimental findings, especially the low field side cooling, are discussed. In a rather restricted parameter range it is shown that ASDEX H-mode plasmas may be operated up to densities at least comparable to the L-mode density limit under similar conditions.}, number={9}, journal={Nuclear Fusion}, author={Stabler. A. et al.}, year={1992}, pages={1557–1583}, language={en} }

@article{Jenko_2000, title={Electron temperature gradient driven turbulence}, volume={7}, ISSN={1070-664X, 1089-7674}, DOI={10.1063/1.874014}, abstractNote={Collisionless electron-temperature-gradient-driven (ETG) turbulence in toroidal geometry is studied via nonlinear numerical simulations. To this aim, two massively parallel, fully gyrokinetic Vlasov codes are used, both including electromagnetic effects. Somewhat surprisingly, and unlike in the analogous case of ion-temperature-gradient-driven (ITG) turbulence, we find that the turbulent electron heat flux is significantly underpredicted by simple mixing length estimates in a certain parameter regime (ŝ∼1, low α). This observation is directly linked to the presence of radially highly elongated vortices (“streamers”) which lead to very effective cross-field transport. The simulations therefore indicate that ETG turbulence is likely to be relevant to magnetic confinement fusion experiments.}, number={5}, journal={Physics of Plasmas}, author={Jenko. F. et al.}, year={2000}, pages={1904–1910}, language={en} }

@article{Zonca_Chen_Santoro_Dong_1998, title={Existence of discrete modes in an unstable shear {A}lfvén continuous spectrum}, volume={40}, ISSN={0741-3335, 1361-6587}, DOI={10.1088/0741-3335/40/12/002}, abstractNote={In this letter, we demonstrate the existence of unstable ion temperature gradient driven Alfve´n eigenmodes in tokamak plasmas, which are ideally stable with respect to magnetohydrodynamics (MHD). Conditions for the destabilization of such modes are quantitatively discussed on the basis of theoretical analyses of the mode dispersion relation, which is given in a compact analytical form. It is emphasized that instability requires both sufﬁciently strong thermal ion temperature gradients and that the plasma be sufﬁciently close to ideal MHD marginal stability.}, number={12}, journal={Plasma Physics and Controlled Fusion}, author={Zonca. F. et al.}, year={1998}, pages={2009–2021}, language={en} }

@article{Heidbrink_2021, title={‘{BAAE}’ instabilities observed without fast ion drive}, volume={61}, ISSN={0029-5515, 1741-4326}, DOI={10.1088/1741-4326/abc4c3}, abstractNote={The instability that was previously identified (Gorelenkov 2009 Phys. Plasmas 16 056107) as a fast-ion driven beta-induced Alfv´en-acoustic eigenmode (BAAE) in DIII-D was misidentified. In a dedicated experiment, low frequency modes (LFMs) with characteristic ‘Christmas light’ patterns of brief instability linked to the safety factor evolution occur in plasmas with electron temperature Te 2.1 keV but modest beta. To isolate the importance of different driving gradients on these modes, the electron cyclotron heating (ECH) power and 80 keV, sub-Alfv´enic neutral beams are altered for 50–100 ms durations in reproducible discharges. Although beta-induced Alfv´en eigenmodes and reversed-shear Alfv´en eigenmodes stabilize when beam injection ceases (as expected for a fast-ion driven instability), the LFMs that were called BAAEs persist. Data mining reveals that characteristic LFM instabilities can occur in discharges with no beam heating but strong ECH. A large database of over 1000 discharges shows that LFMs are only unstable in plasmas with hot electrons but modest overall beta. The experimental LFMs have low frequencies (comparable to diamagnetic drift frequencies) in the plasma frame, occur near the minimum of the safety factor qmin, and appear when qmin is close to rational values. Theoretical analysis suggests that the LFMs are a low frequency reactive instability of predominately Alfv´enic polarization.}, number={1}, journal={Nuclear Fusion}, author={Heidbrink. W.W. et al.}, year={2021}, pages={016029}, language={en} }

@article{Gates_2012, title={Origin of Tokamak Density Limit Scalings}, volume={108}, ISSN={0031-9007, 1079-7114}, DOI={10.1103/PhysRevLett.108.165004}, number={16}, journal={Physical Review Letters}, author={Gates. D. A. et al.}, year={2012}, pages={165004}, language={en} }

@article{Rice_2020, title={Understanding {LOC/SOC} phenomenology in tokamaks}, volume={60}, rights={http://iopscience.iop.org/info/page/text-and-data-mining}, ISSN={0029-5515, 1741-4326}, DOI={10.1088/1741-4326/abac4b}, abstractNote={Phenomenology of Ohmic energy conﬁnement saturation in tokamaks is reviewed. Characteristics of the linear Ohmic conﬁnement (LOC) and saturated Ohmic conﬁnement (SOC) regimes are documented and transformations in all transport channels across the LOC/SOC transition are described, including rotation reversals, “non-local” cut-off and density peaking, in addition to dramatic changes in ﬂuctuation intensity. Uniﬁcation of results from nearly 20 devices indicates that the LOC/SOC transition occurs at a critical value of the product of the density, edge safety factor and device major radius, and that this product increases with toroidal magnetic ﬁeld. Comparison with gyro-kinetic simulations suggests that the effects of sub-dominant TEMs are important in the LOC regime while ITG mode turbulence dominates with SOC.}, number={10}, journal={Nuclear Fusion}, publisher={IOP Publishing}, author={Rice. J.E. et al.}, year={2020}, pages={105001}, language={en} }

@article{Chen_2016_EPL, title={{A}lfvénic ion temperature gradient activities in a weak magnetic shear plasma}, volume={116}, ISSN={0295-5075, 1286-4854}, DOI={10.1209/0295-5075/116/45003}, abstractNote={We report the ﬁrst experimental evidence of Alfv´enic ion temperature gradient (AITG) modes in HL-2A Ohmic plasmas. A group of oscillations with f = 15–40 kHz and n = 3–6 is detected by various diagnostics in high-density Ohmic regimes. They appear in the plasmas with peaked density proﬁles and weak magnetic shear, which indicates that corresponding instabilities are excited by pressure gradients. The time trace of the ﬂuctuation spectrogram can be either a frequency staircase, with diﬀerent modes excited at diﬀerent times or multiple modes may simultaneously coexist. Theoretical analyses by the extended generalized ﬁshbone-like dispersion relation (GFLDR-E) reveal that mode frequencies scale with ion diamagnetic drift frequency and ηi, and they lie in KBM-AITG-BAE frequency ranges. AITG modes are most unstable when the magnetic shear is small in low pressure gradient regions. Numerical solutions of the AITG/KBM equation also shed light on why AITG modes can be unstable for weak shear and low pressure gradients. It is worth emphasizing that these instabilities may be linked to the internal transport barrier (ITB) and H-mode pedestal physics for weak magnetic shear.}, number={4}, journal={Europhysics Letters}, author={Chen.W. et al.}, year={2016},  pages={45003}, language={en} }

@article{Connor_1978, title={Shear, Periodicity, and Plasma Ballooning Modes}, volume={40}, ISSN={0031-9007}, DOI={10.1103/PhysRevLett.40.396}, number={6}, journal={Physical Review Letters}, author={Connor. J. W. et al.}, year={1978}, pages={396–399}, language={en} }

@article{Belli_Candy_2010, title={Fully electromagnetic gyrokinetic eigenmode analysis of high-beta shaped plasmas}, volume={17}, ISSN={1070-664X, 1089-7674}, DOI={10.1063/1.3495976}, abstractNote={A new, more efficient method to compute unstable linear gyrokinetic eigenvalues and eigenvectors has been developed for drift-wave analysis of plasmas with arbitrary flux-surface shape, including both transverse and compressional magnetic perturbations. In high-beta, strongly shaped plasmas like in the National Spherical Torus Experiment (NSTX) [M. Ono et al., Nucl. Fusion 40, 557 (2000)], numerous branches of closely spaced unstable eigenmodes exist. These modes are difficult and time-consuming to adequately resolve with the existing linear initial-value solvers, which are further limited to the most unstable eigenmode. The new method is based on an eigenvalue approach and is an extension of the GYRO code [J. Candy and R. E. Waltz, J. Comput. Phys. 186, 545 (2003)], reusing the existing discretization schemes in both real and velocity-space. Unlike recent methods, which use an iterative solver to compute eigenvalues of the relatively large gyrokinetic response matrix, the present scheme computes the zeros of the much smaller Maxwell dispersion matrix using a direct method. In the present work, the new eigensolver is applied to gyrokinetic stability analysis of a high-beta, NSTX-like plasma. We illustrate the smooth transformation from ion-temperature-gradient (ITG)-like to kinetic-ballooning (KBM)-like modes, and the formation of hybrid ITG/KBM modes, and further demonstrate the existence of high-k Alfvénic drift-wave “cascades” for which the most unstable mode is a higher excited state along the field line. A new compressional electron drift wave, which is driven by a combination of strong beta and pressure gradient, is also identified for the first time. Overall, we find that accurate calculation of stability boundaries and growth rates cannot, in general, ignore the compressional component δB∥ of the perturbation.}, number={11}, journal={Physics of Plasmas}, author={Belli. E. A. et al.}, year={2010}, pages={112314}, language={en} }

@article{Kumar_2021, title={Turbulent transport driven by kinetic ballooning modes in the inner core of {JET} hybrid {H}-modes}, volume={61}, ISSN={0029-5515, 1741-4326}, DOI={10.1088/1741-4326/abd09c}, abstractNote={Turbulent transport in the inner core of the high-β JET hybrid discharge 75225 is investigated extensively through linear and non-linear gyro-kinetic (GK) simulations using the GK code GKW in the local approximation limit. Compared to previous studies (Citrin et al 2015 Plasma Phys. Control. Fusion 57 014032; Garcia et al 2015 Nucl. Fusion 55 053007), the analysis has been extended towards the magnetic axis, ρ < 0.3, where the turbulence characteristics remain an open question. Understanding turbulent transport in this region is crucial to predict core profile peaking that in turn will impact the fusion reactions and the tungsten neoclassical transport, in present devices as well as in ITER. At ρ = 0.15, a linear stability analysis indicates that kinetic ballooning modes (KBMs) dominate, with an extended mode structure in ballooning space due to the low magnetic shear. The sensitivity of KBM stability to main plasma parameters is investigated. In the non-linear regime, the turbulence induced by these KBMs drives a significant ion and electron heat flux. Standard quasi-linear (QL) models are compared to the non-linear results. The standard reduced QL models work well for the E × B fluxes, but fail to capture magnetic flutter contribution to the electron heat flux induced by the non-linear excitation of low kθρi micro-tearing modes that are linearly stable. An extension of the QL models is proposed allowing better capturing the magnetic flutter flux.}, number={3}, journal={Nuclear Fusion}, author={Kumar. N. et al.}, year={2021},  pages={036005}, language={en} }

@article{Xu_Chen_2023, title={Gyrokinetic analysis of turbulent transport by electromagnetic turbulence in finite β plasmas with weak magnetic shear on {HL-2A}}, volume={63}, ISSN={0029-5515, 1741-4326}, DOI={10.1088/1741-4326/acff78}, abstractNote={Turbulent transport and zonal flow (ZF) dynamics in the mixed kinetic ballooning mode (KBM) and ion temperature gradient (ITG) mode dominated turbulence states are analyzed by gyrokinetic simulations based on the experimental observations of KBMs in the HL-2A Internal Transport Barrier (ITB) plasmas with weak magnetic shear configuration. Results have shown that KBMs and ITGs are the main candidates for turbulent transport and the inclusion of fast ions (FIs) can destabilize both instabilities, which is resulted by the decrease of ballooning parameter αMHD hence reduced Shafranov shift stabilization due to the negative FI density gradient, making them more unstable according to the ˆs-αMHD diagram. In addition, the ITGs are suppressed by both dilution and finite-β stabilization effects caused by FIs. Nonlinear simulations excluding the effects of FIs indicate that the transport is minimum at βe = βeexp as the turbulence predominated by ITGs is strongly suppressed by the nonlinear electromagnetic (EM) stabilization and enhancement of ZFs. However, the transport is further increased with βe although the ZFs become stronger due to the transition from ITG to KBM dominated turbulence state. The presence of FIs can modify the relation between ZF shearing rate and βe. The transport level is insensitive to βe when βe ⩽ βeexp but increases significantly because of the KBM destabilization. Meanwhile, the ZF amplitude reaches maximum at βe = βeexp whereas it suffers an erosion at higher values, implying that the plasma might be a self-organized critical state owning to the interactions among ZFs, KBMs and FI effects hence setting the plasma β around βexp. The results have also provided a possible explanation that the destabilization of EM turbulence is responsible for the ion heat transport stiffness under weak magnetic shear configurations in the off-axis neutral beam injection heated ITB plasmas on HL-2A.}, number={12}, journal={Nuclear Fusion}, author={Xu. J. Q. et al.}, year={2023}, pages={126031}, language={en} }

@article{drake1977kinetic,
  title={Kinetic theory of tearing instabilities},
  author={Drake. J.F. et al.},
  journal={The Physics of Fluids},
  volume={20},
  number={8},
  pages={1341--1353},
  year={1977},
  publisher={American Institute of Physics}
}

@article{Zonca_Briguglio_2002, title={Energetic particle mode stability in tokamaks with hollow q -profiles}, volume={9}, ISSN={1070-664X, 1089-7674}, DOI={10.1063/1.1519241}, abstractNote={A thorough analysis of energetic particle modes (EPM) stability and mode structures is presented for tokamaks with hollow q profiles. Focusing on the region near the minimum-q surface, EPM gap modes and resonant EPMs are shown to exist as solutions of the same dispersion relation. By controlling the fast ion distribution function, or, equivalently, their fundamental dynamical properties, a smooth transition between these two classes of modes is obtained within the EPM dispersion relation. When toroidal coupling becomes important, it is demonstrated that EPMs may have either single or double hump radial structures. The local analyses of EPM stability and mode structures near the minimum-q surface are put in the broader framework of EPM stability and EPM induced transport in tokamaks with hollow q profiles and a brief summary is also given of present understanding of such problems based on results of three-dimensional nonlinear hybrid magneto–hydrodynamic–gyrokinetic simulations. Possible implications of present results are discussed in terms of experimental observations and possibilities of designing novel experimental setups to probe, at least conceptually, the complex predictions of theory.}, number={12}, journal={Physics of Plasmas}, author={Zonca. F. et al.}, year={2002}, pages={4939–4956}, language={en} }

@article{Long_2021, title={Enhanced particle transport events approaching the density limit of the J-TEXT tokamak}, volume={61}, ISSN={0029-5515, 1741-4326}, DOI={10.1088/1741-4326/ac36f2}, abstractNote={Enhanced particle transport events are discovered and analyzed as the density limit of the J-TEXT tokamak is approached. Edge shear layer collapse is observed and the ratio of Reynolds power to turbulence production decreases. Simultaneously, the divergence of turbulence internal energy flux (i.e. turbulence spreading) increases, indicating that shear layer collapse triggers an outward spreading event. Studies of correlations show that the enhanced particle transport events are quasi-coherent, and manifested primarily in density fluctuations which exhibit positive skewness. Electron adiabaticity emerges as the critical parameter which signals transport event onset. For α < 0.35 as density approaches the Greenwald density, both turbulence spreading and density fluctuations rise rapidly. Taken together, these results elucidate the connections between edge shear layer, density fluctuations, particle transport events, turbulence spreading and plasma edge cooling as the density limit is approached.}, number={12}, journal={Nuclear Fusion}, author={Long. T. et al.}, year={2021}, pages={126066}, language={en} }

@article{Long_2024, title={The role of shear flow collapse and enhanced turbulence spreading in edge cooling approaching the density limit}, volume={64}, ISSN={0029-5515, 1741-4326}, DOI={10.1088/1741-4326/ad3e15}, abstractNote={Experimental studies of the dynamics of shear flow and turbulence spreading at the edge of tokamak plasmas are reported. Scans of line-averaged density and plasma current are carried out while approaching the Greenwald density limit on the J-TEXT tokamak. In all scans, when the Greenwald fraction fG = ¯n/ nG = ¯n/ (Ip / π a2) increases, a common feature of enhanced turbulence spreading and edge cooling is found. The result suggests that turbulence spreading is a good indicator of edge cooling, indeed better than turbulent particle transport is. The normalized turbulence spreading power increases significantly when the normalized E × B shearing rate decreases. This indicates that turbulence spreading becomes prominent when the shearing rate is weaker than the turbulence scattering rate. The asymmetry between positive/negative (blobs/holes) spreading events, turbulence spreading power and shear flow are discussed. These results elucidate the important effects of interaction between shear flow and turbulence spreading on plasma edge cooling.}, number={6}, journal={Nuclear Fusion}, author={Long. T. et al.}, year={2024}, pages={066011}, language={en} }

@article{Xiaoquan_2006, title={Identification and Analysis of Magnetic Structures on HL-2A}, volume={8}, ISSN={1009-0630}, DOI={10.1088/1009-0630/8/6/04}, abstractNote={A method for the identification and analysis of magnetic islands is presented based on the calculation of the perturbative current and magnetic field in plasmasA. cylindrical approximation is adopted and the toroidal effect on plasma equilibrium is also included. This method has been used on the HL-2A tokamak for analysing the magnetic island structures.}, number={6}, journal={Plasma Science and Technology}, author={X. Q. Ji . et al.}, year={2006}, pages={644–648}, language={en} }

\end{document}